%% file: main_emnlp.tex
\pdfoutput=1
\documentclass[11pt]{article}

\usepackage{acl}

\usepackage[T1]{fontenc}
\usepackage[utf8]{inputenc}
\usepackage{times}
\usepackage{latexsym}
\usepackage{inconsolata}
\usepackage{microtype}

\usepackage{graphicx}

\usepackage{soul}
\usepackage{url}
\usepackage{amsmath,amsthm,amssymb}
\usepackage{booktabs}
\usepackage{algorithm}
\usepackage{algorithmic}
\usepackage{enumitem}
\usepackage[switch]{lineno}

\usepackage{caption}
\usepackage{subcaption}

\usepackage{multicol}
\usepackage{multirow}
\usepackage{xspace}
\usepackage{xcolor}
\usepackage{fancyhdr}
\usepackage{makecell}
\usepackage{pifont}
\usepackage{dsfont}

\usepackage{booktabs,makecell}   % already used in your tables
\newcommand{\cmark}{\ding{51}}%
\newcommand{\xmark}{\ding{55}}

\DeclareMathOperator*{\argmax}{arg\,max}

\title{Do Code Semantics Help? A Comprehensive Study on Execution Trace-Based Information for Code Large Language Models}

\author{
  \textbf{Jian Wang\textsuperscript{1,2}}, 
  \textbf{Xiaofei Xie\textsuperscript{1}}, 
  \textbf{Qiang Hu\textsuperscript{3}\thanks{Corresponding Author: qianghu@tju.edu.cn}}, 
  \textbf{Shangqing Liu\textsuperscript{4}}, 
  \textbf{Yi Li\textsuperscript{2}} \\[1ex]
  \textsuperscript{1}Singapore Management University, Singapore \\
  \textsuperscript{2}Nanyang Technological University, Singapore \\
  \textsuperscript{3}Tianjin University, China\\
  \textsuperscript{4}State Key Laboratory for Novel Software Technology, Nanjing University, China \\[1ex]
}

\begin{document}

\maketitle
\cfoot{\thepage}
\renewcommand{\headrulewidth}{0pt} 
\renewcommand{\footrulewidth}{0pt}

\begin{abstract}
Code Large Language Models (Code LLMs) have opened a new era in programming with their impressive capabilities. However, recent research has revealed critical limitations in their ability to reason about runtime behavior and understand the actual functionality of programs, which poses significant challenges for their post-training and practical deployment. Specifically, Code LLMs encounter two principal issues: (1) a lack of proficiency in reasoning about program execution behavior, as they struggle to interpret what programs actually do during runtime, and (2) the inconsistent and fragmented representation of semantic information, such as execution traces, across existing methods, which hinders their ability to generalize and reason effectively. These challenges underscore the necessity for more systematic approaches to enhance the reasoning capabilities of Code LLMs. To address these issues, we introduce a generic framework to support integrating semantic information~(e.g., execution trace) to code task-relevant prompts, and conduct a comprehensive study to explore the role of semantic information in enhancing the reasoning ability of Code LLMs accordingly. Specifically, we focus on investigating the usefulness of trace-based semantic information in boosting supervised fine-tuning~(SFT) and post-phase inference of Code LLMs. The experimental results surprisingly disagree with previous works and demonstrate that semantic information has limited usefulness for SFT and test time scaling of Code LLM. 

% on various code tasks demonstrate that semantic information has limited help for SFT which is different from previous research findings. However, accompanied by test-time scaling, semantic information significantly improves the performance of Code LLMs by up to 10.85.

% Our contributions are as follows: (1) We introduce and opensource a trace-based framework to enhance the training and inference of Code Large Language Models (LLMs), enabling them to acquire semantic-related knowledge and improve their understanding of program behavior. (2) We curate five execution representation techniques from prior work and systematically evaluate their effectiveness during fine-tuning across three LLMs. (3) Additionally, we conduct an empirical study on scaling inference, incorporating feedback mechanisms, reward models, and trace representations during test-scaling computations, revealing that reasoning in the inference phase significantly improves effectiveness. (4) Finally, we construct and opensource a high-quality dataset supporting multiple representation formats, facilitating robust evaluation and fine-tuning of Code LLMs.

\end{abstract}

\section{Introduction}

\input{sections/p1_intro_related_works}

\section{Problem Statement}

\input{sections/problem_state}

\section{Evaluation Framework}

\input{sections/method}

\section{Experiment Design}

\input{sections/p4_experiments}

\section{Result Analysis}

\input{sections/results}

\section{Related works}

\input{sections/related_works}

\section{Conclusion}
\input{sections/p5_conclusion}

\section{Data and Source Code availability}
All source code, datasets, and intermediate data for reproduction are available at ~\url{https://github.com/tracewise-probing/tracewise_probing}.

\section{Acknowledgements}
This work was partially supported by the National Research Foundation, Singapore, and the Cyber Security Agency under its National Cybersecurity R\&D Programme (NCRP25-P04-TAICeN), the CyberSG R\&D Cyber Research Programme Office, the Singapore Ministry of Education Academic Research Fund Tier 1 (RG12/23). Any opinions, findings, and conclusions or recommendations expressed in this material are those of the author(s) and do not necessarily reflect the views of National Research Foundation, Singapore, Cyber Security Agency of Singapore, CyberSG R\&D Programme Office as well as MOE.

\clearpage
\newpage
% While the majority of our experiments were conducted on A100 GPUs, we believe our findings and conclusions are broadly compatible with other GPU types, such as H100, A6000, and RTX 4090, BFloat64 supported.

% \section*{Limitations}
% \section*{Ethical Considerations}

\clearpage
\input{sections/limitation}
\bibliography{custom}

\clearpage
\newpage 
\appendix

\input{sections/p6_appendix_1}

\input{sections/p6_appendix_2}

\input{sections/p6_appendix_3}

\end{document}

%% file: sections/p1_intro_related_works.tex
Code large language models (Code LLMs) have emerged as prominent programming assistants, demonstrating remarkable performance across various coding tasks, including program repair~\cite{conversationalrepair}, code generation~\cite{evalplus}, and code summarization~\cite{codesearchnet}. Recently, several Code LLMs have been introduced, each characterized by distinct training schemes. For instance, Llama3.1~\cite{llama31,codellama}, is fine-tuned with code infilling tasks and long code input contexts, complemented by an instruction fine-tuning process. Similarly, DeepSeek-Coder~\cite{guo2024deepseek} is trained on more than 2 trillion tokens using a fill-in-the-blank task to enhance its code generation capabilities. These models focus on learning contextual information from code and docstrings, advancing their general understanding of code~\cite{chen2024evaluating,ni2024next,ding2024semcoder}.

However, these approaches predominantly capture the static dimensions of code~(e.g., tokens and context), while neglecting the dynamic semantics crucial for a comprehensive understanding of code. Recent studies have highlighted this limitation, revealing that even state-of-the-art models like GPT-4 struggle to reason about runtime behaviors of code~\cite{chen2024evaluating}. Understanding code semantics and accurately predicting runtime behavior is critical, particularly for practical coding applications that require semantic understanding~\cite{ni2024next,ding2024semcoder}. This underscores the urgent need to investigate methodologies that enhance the reasoning capabilities of Code LLMs for better semantic understanding, such as coverage prediction~\cite{chen2024evaluating} and output prediction~\cite{chen2024evaluating,codeexecutor} abilities.

To address these challenges, two primary methods have been proposed to bolster the reasoning capabilities of Code LLMs in code generation tasks: 1) \textit{inference with semantic information} -- integrating feedback, such as error information to the inputs~(or outputs) at inference time, and 2) \textit{fine-tuning with semantic information} -- directly enhancing the models' intrinsic reasoning by integrating semantically enriched fine-tuning data, thereby enabling improved predictions~\cite{ni2024next,ding2024semcoder}. In this work, we consider both approaches and explore whether and how incorporating code semantic information can enhance the performance of Code LLMs. Specifically, we investigate evaluate their effectiveness across various code-related tasks.

The core challenge lies in identifying and collecting appropriate semantic data for training or inference to improve the semantic comprehension of Code LLMs. Recent efforts~\cite{ding2024semcoder,ni2024next} have begun to address this by fine-tuning models using dynamic data, such as execution behaviors, to enhance semantic understanding. While these approaches have shown promise, they employ diverse semantic representations, such as natural language descriptions of programs or execution traces. There remains a lack of 1) a unified study and tool that supports all semantic representations, and 2) systematic understanding regarding how different training data compositions, particularly in terms of semantic representations, impact code reasoning and generation capabilities.

% This paper conducts a comprehensive study on enhancing reasoning and semantic understanding in \textbf{multiple software engineering benchmarks, like code generation, code reasoning and code repair tasks through fine-tuning, inference with feedback and Scaling Inference in test-computation}. However, several challenges complicate this investigation: 1) the composition of training data pairs $(X, Y)$, where the choice of inputs (i.e., $X$) and outputs (i.e., $Y$) can vary significantly, such as using natural language descriptions or code execution traces; 2) the representation of execution behavior, which critically influences the effectiveness of trained code models~\cite{feng2020codebert,guo2020graphcodebert,ni2024next,ding2024semcoder}; and 3) the absence of high-fidelity datasets that provide diverse semantic representations. Existing studies often generate data using LLMs~\cite{ding2024semcoder,ni2024next}, which may not always provide accurate information, and these datasets are typically not publicly available~\footnote{To the best of our knowledge, SemCoder only provides trained data, excluding the curating script.}.

% Scaling LLM Test-Time Compute Optimally can be More Effective than Scaling Model Parameters

To this end, we propose and implement a generic framework that facilitates the generation of multiple types of semantic representations, supporting post-training, one-time inference, and the scaling of test-time computation during inference.  Based on this framework, we conduct a systematic study to explore the efficacy of semantic information in boosting code generation. Specifically, we integrate different semantic representations~(i.e., different code execution traces) into the input data~(prompt) during both SFT time and inference time to assess their impact. Additionally, we examine the influence of different training strategies (e.g., parameter-efficient fine-tuning) on the effectiveness of semantic information. Different from previous research findings~\cite{ni2024next,ding2024semcoder}, our experimental results demonstrate that integrating existing semantic information into the input prompt has limited benefits to the performance of tuned Code LLMs. 
% Inspired by the critical role of debugging and refinement in code generation, as highlighted in recent studies~\cite{chen2023teaching,jiang2024training,xia2023keep}, we draw parallels to human coding practices, where code is often refined through internal reasoning processes. We hypothesize that a repair-based framework can similarly enhance the reasoning capabilities of LLMs in code generation. For the repair process, the input $X$ and output $Y$ must at least include the buggy code and its corresponding patch. Additionally, semantic data such as runtime information (e.g., execution traces) and code specifications (e.g., descriptions) can be incorporated into both $X$ and $Y$ to further improve learning. To evaluate their effectiveness, we integrate three execution trace representations from recent works—NExT~\cite{ni2024next}, Code Executor~\cite{codeexecutor}, and SemCoder~\cite{ding2024semcoder}—into our study. By analyzing their limitations, we develop a novel representation that is simpler and more effectively learned by Code LLMs. \wj{Our experimental results demonstrate its superior performance compared to existing representations.}

% Finally, to provide a comprehensive training dataset encompassing multiple semantic layers, we construct a new dataset featuring bug-patch function pairs along with unit tests. This dataset is designed to be versatile, offering a wealth of semantic data across diverse representations. Our contributions are as follows:

To summarize, the main contributions of this paper are:
% \wj{The contribution: Emp-study on finetune and scaling inference test-time, Our framework is opensource, a high-quality dataset.  }

\begin{itemize}[leftmargin=*]
    \item We introduce the first generic framework that supports different types of code semantic representations. Based on this framework, we construct and open-source a high-fidelity dataset featuring diverse execution behavior representations, including bug-patch function pairs, unit tests, and multiple semantic layers. The dataset and all related implementations are publicly available on our website \footnote{\url{https://github.com/tracewise-probing/tracewise_probing}}.
    \item We conduct a comprehensive study to explore the effectiveness of semantic information in enhancing both SFT and inference of code LLMs.
    \item We summarize multiple findings, such as semantic information integrated in the input does not positively contribute to the inference.

\end{itemize}

% Through extensive experiments, we demonstrate that our approach significantly improves the reasoning and generation capabilities of Code LLMs, paving the way for more effective programming assistants.

% The rest of this paper is organized as follows. \todo{XXX}

%% file: sections/problem_state.tex
% \subsection{Problem Definition}
\label{sec:2_1}

We address the problem of enhancing the code generation capabilities of Code LLMs by integrating code semantic information into the input prompts. Current Code LLMs primarily rely on static text data, which often fails to capture the nuanced semantics crucial for thorough code understanding~\cite{wei2023magicoder,wei2024selfcodealign,abdin2024phi}. Inspired by the practices of human developers, who iteratively refine code through reasoning and semantic assessment rather than relying solely on dynamic testing or runtime feedback, we note that such reasoning and refinement processes are largely absent in existing models~\cite{ni2024next,ding2024semcoder}.  There are two main sub-problems we are interested in for Code LLMs, 1) fine-tuning with semantic information, and 2) inference with semantic information.

% \wj{1) fine-tuning with semantic information in small LLM, size not exceed 7B. and 2) inference with semantic information in larger LLM or close-source commercial LLM}

% (1) Fine-tuning with semantic information. Our Code LLM fine-tuning paradigm focuses on a repair-based fine-tuning framework that leverages a carefully curated dataset $\mathcal{D}=\{(x, y)\}$, where \(x\) and \(y\) represent the inputs and outputs of a Code LLM \(p_\theta\). Specifically, The dataset $\mathcal{D} =\Bigl\{\,(x,y)\;\big|\; x=(b,\,r_b),\; y=(r_a,\,a) $ curated includes pairs of buggy code \(b\), its corresponding capture semantics~(execution traces) \(r\), and corresponding answer \(a\) to patched code, respectively, and \(r\) is optional. A central question is how to optimally represent reasoning information in the training data to improve the model's performance.

% * $b$: buggy code  
% * $r_b$: rationale/trace for $b$ (always present)  
% * $a$: patched code  
% * $r_a$: optional rationale for $a$
%    If $r_a=\varnothing$, then $y=(a)$ and the summation in (2) is over the tokens of $a$ alone.

% Given a dataset $D$, the optimzation to policy $p_{\theta}$ is given by:

(1) Fine-tuning with semantic information. Our Code LLM fine-tuning paradigm focuses on a \textit{repair-based} fine-tuning framework that leverages a carefully curated dataset $\mathcal{D} \;=\; \bigl\{\, (x,y) \;\big|\; x=(b,r),\; y=a \bigr\}$ where (x, y) is the input-output pair for model fine-tuning, \(b\) denotes the buggy code fragment, \(r\) is execution-trace rationale, and \(a\) represents the patched code. We selected the code repair task because models often fail to produce correct code in a single attempt, requiring the refinement of the output until it is correct.
A central question is how to encode the reasoning signals $r$ so that the Code LLM can learn better code generation capability. 
% % \(b\) denotes the buggy code fragment, \(r_b\) its execution-trace rationale (always present), the corresponding answer \(a\) to patched code, and \(r_a\) an optional rationale accompanying the patch.  
% If \(r_b=\varnothing\) and \(r_a=\varnothing\), this simplifies the training of LLM in generation capability instead of reasoning and generation of both.   A central question is how to encode the reasoning signals \((r_b,r_a)\) so that the policy \(p_{\theta}\) learns to generate higher quality fixes.

% \begin{equation}
% \mathcal{L}(\theta)=
% -\;
% \mathbb{E}_{(x,y)\sim\mathcal{D}}
% \Bigl[
% \sum_{t=1}^{|y|}
% \log p_{\theta}\!\bigl(y_t \mid x,\,y_{<t}\bigr)
% \Bigr]
% \end{equation}

% \qiang{check here}\wj{done}
(2) Scaling Inference through Semantic Refinement. Emulating human “work-and-check” practices at inference time—iteratively refining candidate solutions and verifying each step—can substantially improve an LLM’s accuracy under a fixed but non-trivial test-time compute budget \(N\)~\cite{sky_t1_2025}. The key question is: \textit{Given a fixed inference-time compute budget \(N\), to what extent can an LLM improve its performance when prompts are enhanced with semantic representations?} In this paradigm, a search-based computation strategy \(\theta\)  specifies how to (1) propose candidate solutions incrementally, (2) verify or score each partial output (e.g., via code execution or a reward model), and (3) refine solutions based on feedback—all within the budget \(N\). Formally, following \cite{snell2024scaling}, for a given query \(q\), the final output \(y\) is drawn from

\begin{equation}
y \;\sim\; \operatorname{Target}\bigl(\theta,\; q,\; N,\; \mathrm{Verify}\bigr)
\end{equation}

where \(\operatorname{Target}(\theta, q, N, \mathrm{Verify})\) is a test-scaling framework which iterates through proposing, verifying, pruning, and refining partial solutions until the budget \(N\) is exhausted. If \(y^*(q)\) denotes the ground-truth correct answer for \(q\), we measure accuracy via the indicator \(\mathds{1}_{\{y = y^*(q)\}}\). 
% In general, we want to optimize our use of compute on each problem \(q\), 
 In general, we define the test-time compute-optimal strategy \(\theta_{q}^*(N)\) as the one that maximizes the expected probability of generating the correct answer:

% \begin{equation}
% \theta_{q, a^*(q)}^*(N)=\operatorname{argmax}_\theta\left(\mathbb{E}_{y \sim \operatorname{Target}(\theta, N, q, \mathrm{Verify})}\left[\mathds{1}_{y=y^*(q)}\right]\right)
% \end{equation}
\begin{multline}
\theta_{q}^*(N)
= \argmax_\theta \Bigl(
   \mathbb{E}_{y \sim \operatorname{Target}(\theta, N, q, \mathrm{Verify})} \\[-0.5ex]
   \bigl[\mathds{1}_{y = y^*(q)}\bigr]
\Bigr)
\end{multline}

Here, \(\theta\) may control how many refinement steps to run, which candidate paths to verify, subject to the budget \(N\). Verification signals (such as code execution) enable the model to discard incorrect paths or improve partially correct ones, while iterative refinement uses feedback to converge on better outputs. By strategically allocating test-time resources, a trace-based verify-and-refine loop can substantially boost accuracy without additional training.

%% file: sections/method.tex
% \textbf{Other different strategies } \\

% \textit{Best-of-N Sampling (BoN)} is a simple yet effective search method that functions as a solution-level tree search, where multiple candidate solutions are generated by a model, and the best solution is selected using a reward model while discarding the rest~\cite{cobbe2021training}. When oracle rewards—such as comparisons with ground truth—are available, BoN's accuracy improves with more samples due to increased coverage~\cite{brown2024large}. However, in most practical scenarios, oracle rewards are inaccessible, making the learning of an accurate reward model the primary bottleneck. The effectiveness of BoN thus heavily depends on the quality of the reward model, highlighting the need for advancements in reward model learning to fully leverage this approach.

% \includegraphics[width=0.8\textwidth]{sections/figures/inference_scaling.pdf}

% \subsubsection{Enhancing Reasoning via Code Refinement}\label{sec:repairrationale}
\begin{figure}[t]
    \small
	\centering
	\includegraphics[width=.95\linewidth]{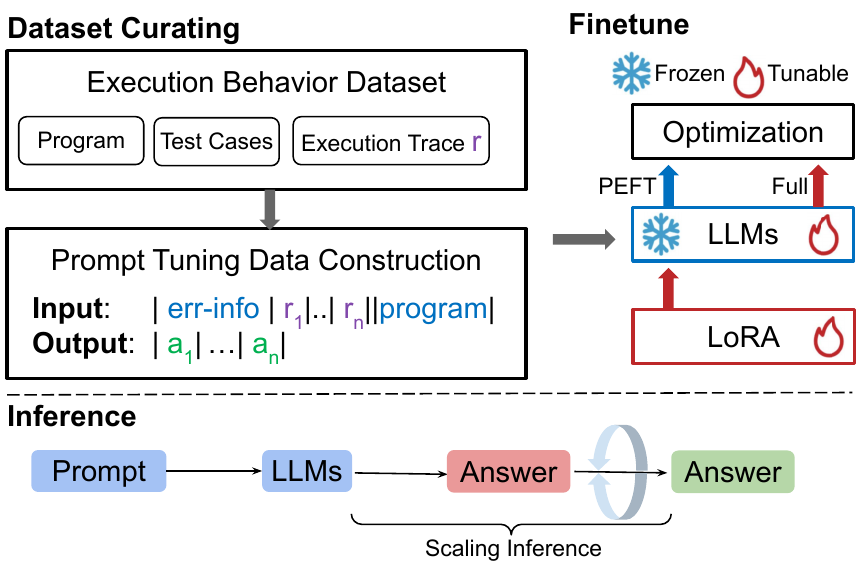}
	\caption{Paradigm of our framework. Initially, we curate prompt-tuning data from the Execution Behavior Dataset by extracting runtime execution messages, which are then formalized using a trace adapter. Subsequently, we employ parameter-efficient fine-tuning techniques, such as LoRA, or opt for full parameter fine-tuning to train the foundation model. In the output above, the purple text denotes the rationale, while the green text represents the answer. During the inference phase, the framework supports Scaling Inference to enhance the capability of LLMs.}
	\label{fig:overview}
\end{figure}

% \begin{figure*}[t]
%     \small
% 	\centering
% 	\includegraphics[width=0.95\linewidth]{sections/figures/inference_scaling.pdf}
% 	\caption{Scaling inference at test time with different search strategies,[Snell et al., 2024].\qiang{consider to remove this figure}}
% 	\label{fig:scaling}
% \end{figure*}

% The outputs of LLMs are highlighted with \underline{underline}
\begin{figure*}[t]
    \small
	\centering
	\includegraphics[width=1\linewidth]{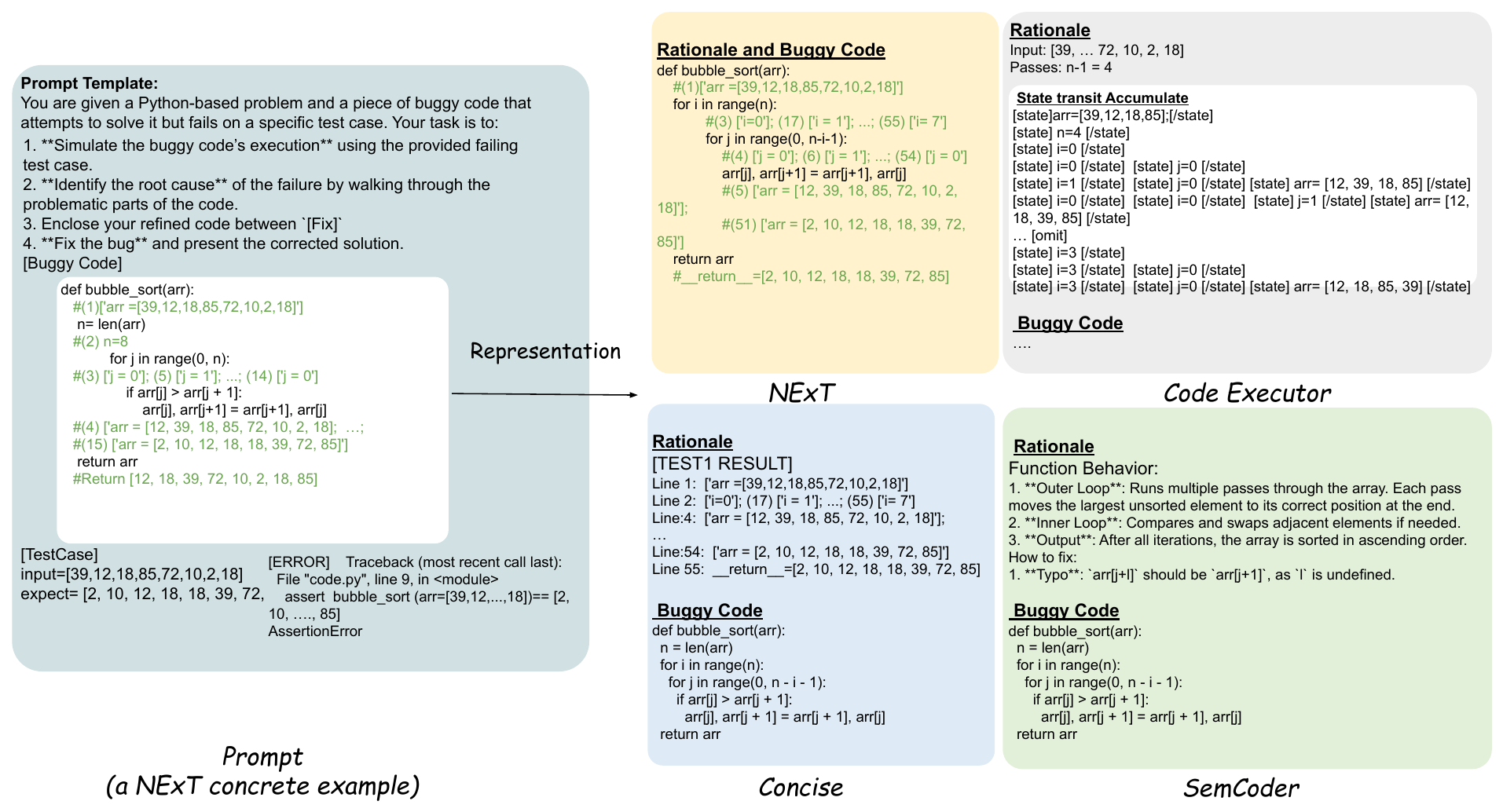}
	\caption{A concrete prompt example~(left panel) and examples of different semantic representations~(right panel).    
    }
	\label{fig:trace}
\end{figure*}

% In complex tasks, scaling inference at test time can significantly boost performance. The figure compares various search strategies (from left to right): Best-of-N (BoN): Generates N complete solutions and then selects the best one according to the reward model’s final score. Beam Search: Expands N candidates at each step, ranks them based on the reward model, and retains the top M candidates for continued exploration. Sequential Revision: Uses external tools to identify the strongest candidate within each iterative chain, then incorporates feedback into subsequent iterations to refine the final answer. Greedy selects the highest-probability token based on the model’s own predicted distribution. This image inspired by [Snell et al.,2024]

% https://docs.google.com/drawings/d/1em97E92oBQP9y-9t7oMKC3mhuZbCKMBOMwMLvnLvyZU/edit?usp=sharing
% This image inspired by~\cite{snell2024scaling}
% % \subsubsection{Enhancing Reasoning via Code Refinement}\label{sec:repairrationale}
% \begin{figure}[t]
%     \small
% 	\centering
% 	\includegraphics[width=\linewidth]{sections/figures/PT_old.pdf}
% 	\caption{\wj{replace later, it is a placeholder picture }.}
% 	\label{fig:overview}
% \end{figure}

\subsection{Overall Design}
This work aims to investigate the impact of various execution-trace based semantics and their representations, denoted as $r$, on the performance of Code LLMs. Following recent works~\cite{ding2024semcoder,chen2024evaluating,ni2024next}, we consider high-level program descriptions and low-level execution traces as potential semantic components of $r$. Different trace representations can significantly influence fine-tuning and inference performance, prompting us to explore which representations best enhance Code LLMs. To achieve the goal, we design and build a framework for the generation and evaluation of automatic semantic representations, which is outlined in Figure~\ref{fig:overview}. The framework consists of three main components, data construction, Code LLM fine-tuning, and inference. 

% \qiang{better to follow the figure to introduce the framework, execution behavior dataset construction, prompt tuning data construction, fine-tuning, and inference.}

\textbf{Fine-Tuning Data Construction.}\label{sec:dataconstruct} Our dataset contains two parts, a program repair dataset and other downstream datasets. The program repair dataset is used to help Code LLMs learn semantic information that is integrated into the buggy/correct code pairs. Specifically, the input $b$ includes \textit{program descriptions}, \textit{test cases}, which are associated with the buggy code  (e.g., an instruction to generate code, a test case fails on the buggy code), and semantic information $r$. For $r$, our framework automatically generates \textit{execution traces} as self-explanations for bug fixes (i.e., patch code $a$) based on the Trace Adapter component. Specifically, Trace Adapter first runs the code using a compiler to collect the raw execution trace. After that, it transfers the raw trace to various trace representations. Currently, Trace Adapter supports five types of representative reasoning-based code semantic information,  Scratchpad~\cite{Scratchpads}, NExT~\cite{ni2024next}, SemCoder~\cite{ding2024semcoder}, CodeExecutor~\cite{codeexecutor}), and Concise which is a new variant of CodeExecutor designed by us. Based on our data construction pipeline, we prepare and open-source a new dataset encompassing buggy/patched code, unit tests, program descriptions, and various trace types, as existing datasets~\cite{ni2024next,ding2024semcoder} do not meet these requirements.

% For Input $b$, we include \textit{program descriptions}, \textit{test cases}, which are associated with the buggy code  (e.g., instruction to generate code, a test case fails on the buggy code). For $r$, our framework leverages \textit{execution traces} as self-explanations for bug fixes (i.e., patch code $a$). Specifically, for the trace representation curated by Trace adapters, we explore most existing ones (i.e., Scratchpad~\cite{Scratchpads}, NExT~\cite{ni2024next}, SemCoder~\cite{ding2024semcoder}, and CodeExecutor~\cite{codeexecutor}) and introduce a novel representation that is simpler and more effectively learned by Code LLMs. \wj{add scaling inference }

\textbf{Supervised Fine-Tuning.}  Our framework supports a two-stage Code LLM fine-tuning process. Concretely, it first fine-tunes Code LLMs using a repair-based workflow~(as introduced in Section~\ref{sec:2_1}) to force models to learn the semantic information hidden in the difference between the execution traces of the buggy and correct codes. Then, other downstream task datasets, such as code generation datasets are used in the second phase to help Code LLMs learn domain-specific knowledge.

% we fine-tune three state-of-the-art Code LLMs with diverse combinations of training data. Our experiments are designed to address the following research questions: (1) the impact of execution trace representation on code generation tasks, code repair, and code reasoning; (2) the effectiveness of train strategy, full parameters tuning, and LoRA-based enhancing reasoning capabilities; (3) the influence of trace representation on code reasoning when scaling inference's test-time computation. (4) and the usefulness of including such representation in enhancing the certainty of LLM. Through this comprehensive study, we aim to provide valuable insights into optimizing the reasoning and generation capabilities of Code LLMs.

\textbf{Test Time Scaling.} Our framework supports two test time scaling strategies, Sequential Scaling and Parallel Scaling ~\cite{khattab2024dspy,sky_t1_2025,wang2025m1,shi2025heimdall}.

\textit{Sequential Scaling} iteratively generates outputs based on the feedback from the previous round. Given an input prompt, the Code LLM first samples $N$ candidate programs and executes each with an external checker (e.g.\ a Python interpreter or trace‐format adapter). If any candidate passes all public test cases, that program is returned immediately; otherwise, the checker emits trace-based diagnostics for every failing candidate. These diagnostics are added to the prompt, prompting the LLM to generate a new revised candidate in the next round. This self-debug~\cite{chen2023teaching} cycle repeats until a correct solution is found or a predefined budget of $R_{\max}$ rounds is reached, exploring at most $N \times R_{\max}$ candidate programs while continuously steering the model with execution-trace feedback. Different from \textit{Sequential Scaling}, \textit{Parallel Scaling} generates multiple solutions at once and selects one accordingly. More implementation details can be found in the Appendix~\ref{exp_config}.

\subsection{Trace Representation Adapters }
The key component in the framework is the trace adapter. Execution traces can be represented in various ways. Our adapter supports various distinct execution representations collected from existing works. Figure~\ref{fig:trace} illustrates examples of each execution representation.

\textbf{NExT} integrates execution traces directly within the code as inline comments. It identifies variables present in each line of code and appends changes in these variables as comments following the respective line, providing a seamless integration of code and state information.

\textbf{SemCoder} utilizes natural language to describe execution traces. It provides a line-by-line explanation of code execution, including aspects such as execution status, variable changes, and input-output relationships. For instance, as shown in Figure~\ref{fig:trace}, SemCoder describes the function signature of `bubble\_sort' and specifies that the `arr' argument accepts only a list of integers, offering a detailed, human-readable explanation.

\textbf{Code Executor} records the state changes of variables in each line, similar to NExT, but presents these execution traces separately from the code, emphasizing a clear distinction between code and execution states.

\textbf{Concise} is a variant of Code Executor, which records the value changes of variables line-by-line and presents the trace separately from the code context as shown in Figure~\ref{fig:trace}. Unlike Code Executor, Concise ignores variables whose values remain unchanged during the execution of a specific line, simplifying the representation. For example, in line 4, the variable `n=10' is omitted in Concise.

%% file: sections/p4_experiments.tex
\begin{table*}[ht]
\centering
\resizebox{0.95\textwidth}{!}
{
\begin{tabular}{c|c|cccccc|cc}
\toprule
\multirow{3}{*}{\textbf{Fine-Tuning}} & \multirow{2}{*}{Task} & \multicolumn{6}{c|}{Program Repair (27.8K samples per each representation)} & \multicolumn{1}{l|}{Code Synthesis} & \multicolumn{1}{l}{Code Reasoning} \\
 &  & Concise & CodeExecutor & NExT & SemCoder(GPT4o) & SemCoder & w/o trace & \multicolumn{1}{c|}{32.4K samples} & 32.4K samples \\
 & Token Size (M) & 23.3 & 23.8 & 19.6 & 33.4 & 32.0 & 12.5 & \multicolumn{1}{c|}{14.4} & 27.8 \\ \hline
\multirow{3}{*}{\textbf{Evaluation}} & \multirow{2}{*}{Task} & \multicolumn{4}{c|}{Code Synthesis} & \multicolumn{2}{c|}{Program Repair} & \multicolumn{2}{c}{Code Reasoning} \\
 &  & HumanEval & MBPP & LiveCodeBench(easy) & \multicolumn{1}{c|}{BigcodeBench(full)} & huamnevalpack(HE-R) & MBPP-R & CRUXEval-I & CRUXEval-O \\
 & Sample Size & 64 & 378 & 880 & \multicolumn{1}{c|}{1140/148} & 164 & 378 & 800 & 800 \\ 
 \bottomrule
\end{tabular}
}
\caption{Details of datasets used in our study.}
\label{tbl_dttrain}
\end{table*}

Based on our framework, we conduct a comprehensive study to explore the usefulness of code semantic information for Code LLMs during fine-tuning and test-time scaling, respectively. Specifically, for the fine-tuning part, we utilize our constructed datasets to fine-tune Code LLMs first, and then evaluate their capabilities on different programming tasks using the basic evaluation paradigm. For the test-time scaling evaluation, we employ different test-time scaling strategies to help assess the ability of fine-tuned Code LLMs from the previous step to investigate the usefulness of semantic information. 

% \subsection{Dataset and Models} 

% \paragraph{Evaluation on Code Generation and Reasoning Tasks}

\textbf{Datasets.} Table~\ref{tbl_dttrain} summarizes datasets used in our study. As fine-tuning contains two stages, it requires two types of datasets. For the first stage, we utilize the framework to prepare different types of semantic information-covered datasets. For the second stage, we use the datasets provided by Semcoder~\cite{ding2024semcoder} for the fine-tuning of downstream tasks.  Regarding the evaluation, we employ widely used datasets to assess the capability of Code LLMs, including Code-Synthesis tasks (HumanEval~\cite{chen2021evaluating}, MBPP~\cite{evalplus}, LiveCodeBench(LCB)~\cite{jain2024livecodebench}, BigcodeBench~\cite{zhuo2024bigcodebench}), two repair tasks (HE-R~\cite{humanevalpack} and MBPP-R collected by us from EvalPlus's MBPP release, regarding their test-failure generation as a source of buggy code.), and two reasoning tasks (CRUXEval-I and CRUXEval-O)~\cite{gu2024cruxeval}.  For the test-time scaling evaluation, we use LiveCodeBench in our experiments, following \cite{sky_t1_2025}. We adapt different trace representations into the prompts while keeping the same dataset and codebase.

% ~\wj{For the test-time scaling study we use the \textbf{LiveCodeBench} (\emph{easy}) split. \textbf{Pass@\!1} is computed on the benchmark’s hidden (private) test cases. Specifically, In \emph{Sequential Revision} we choose the final answer from candidates validated on the benchmark’s public tests, whereas in \emph{Parallel Major-Voting} we select from candidates scored against model-generated test cases rather than the public set.}

\textbf{Models.} For fine-tuning, our study considers three representative LLMs: DeepSeek-Coder~(deepseek-6.7b-base), LLaMA~(Llama3.1-8B), and Gemma2~(gemma2-9b). For the inference, we cover two more closed-source models, GPT-4o and Deepseek-Chat(V3). In addition, for the evaluation of reasoning ability, we include two more models oriented to reasoning, microsoft/phi-4~\cite{abdin2024phi} and AIDC-AI/Marco-o1~\cite{zhao2024marcoo1openreasoningmodels}.

\textbf{Configuration.} Input prompts are produced automatically by the \textsc{DSPy} framework \cite{khattab2024dspy}; full templates can be found in Appendix~\ref{full_example}. All code executes inside a sandbox following the safety procedures of \cite{chen2021evaluating} to guard against malicious generations. For detailed experimental configurations, please refer to Appendix~\ref{exp_config}. Besides, we put the results of HumanEval and HE-R in Appendix~\ref{more_rq1}~\ref{more_rq3} due to the page limitation.

%% file: sections/results.tex
% Our study aims to address the following research questions:

% \paragraph{ \textit{RQ1}:} How effective is different trace representations?
% \paragraph{ \textit{RQ2}:} How effective is different train strageties?
% \paragraph{ \textit{RQ3}:} Given a fixed yet non-trivial inference-time compute budget, to what extent can an LLM improve its performance when presented with prompts derived from different trace representations?
% \paragraph{ \textit{RQ4}:} How useful are self-explanations in $R_Y$ for improving the performance of Code LLMs?

% Through these experiments, we aim to provide a comprehensive understanding of how different components of the training dataset contribute to the performance of Code LLMs and identify the optimal configurations for enhancing their reasoning and generation capabilities.

% \subsection{Execution Trace Representation}\label{sec:tracerep}

\subsection{Fine-Tuning with Semantic Information}

\begin{table*}[t]
\centering
\resizebox{0.95\textwidth}{!}
{

% Preview source code for paragraph 0
\begin{tabular}{c|l|cc|c|ccc|cc}
\toprule
\multicolumn{1}{c}{\multirow{2}{*}{BaseModel}} & \multirow{2}{*}{TrainCorpus} & \multicolumn{2}{c|}{Finetune} & Code Repair & \multicolumn{3}{c|}{Code Synthesis} & \multicolumn{2}{c}{Code Reasoning}\tabularnewline
 &  & downstream & trace & MBPP-R & MBPP & BigcodeBench & LiveCodeBench & CRUXEval-I & CRUXEval-O\tabularnewline
\toprule 
\multirow{9}{*}{DeepSeek-Coder} & - & \textcolor{red}{\xmark} & \textcolor{red}{\xmark} & 17.7 & 71.9 & 41.5 & \underline{40.8} & 40.0 & 40.4\tabularnewline
 & only NL2Code & \textcolor{green}{\cmark} & \textcolor{red}{\xmark} & 25.4 & 72.9 & 43.7 & 12.6 & 60.1 & 55.4\tabularnewline
 & w/o trace & \textcolor{green}{\cmark} &\textcolor{red}{\xmark}  & 39.2 & 75.9 & 45.4 & 35.7 & 61.9 & 56.6\tabularnewline
 & Concise & \textcolor{green}{\cmark} & \textcolor{green}{\cmark} & 39.2 & 74.4 & 44.3 & 29.4 & 61.6 & 55.0\tabularnewline
 & CodeExecutor & \textcolor{green}{\cmark} & \textcolor{green}{\cmark} & 38.4 & \underline{77.2} & 44.6 & 31.5 & 60.4 & 56.1\tabularnewline
 & NeXT & \textcolor{green}{\cmark} &\textcolor{green}{\cmark}  & 37.6 & 76.7 & 44.0 & 36.1 & 61.3 & 54.2\tabularnewline
 % & \ourtrace & \textcolor{green}{\cmark} & \textcolor{green}{\cmark} & 38.9 & \underline{76.7} & 43.9 & 35.7 & 60.0 & 55.6\tabularnewline
 & SemCoder(GPT4o) & \textcolor{green}{\cmark} & \textcolor{green}{\cmark} & 37.0 & 75.7 & 45.4 & 31.5 & \underline{62.0} & \underline{58.1}\tabularnewline
 & SemCoder &  \textcolor{green}{\cmark} & \textcolor{green}{\cmark} & \underline{40.5} & 76.4 & \underline{45.7} & 29.0 & 59.5 & 55.4\tabularnewline
\midrule 
\multirow{9}{*}{LLaMA} & - &  \textcolor{red}{\xmark}&  \textcolor{red}{\xmark}& 20.1 & 58.6 & 31.4 & \underline{27.3} & 42.6 & 36.2\tabularnewline
 & only NL2Code &  \textcolor{green}{\cmark} & \textcolor{red}{\xmark}   & 24.9 & \underline{73.7} & \underline{44.1} & 18.1 & \underline{60.1} & 55.9\tabularnewline
 & w/o trace &  \textcolor{green}{\cmark}  & \textcolor{red}{\xmark}   & 29.1 & 59.1 & 31.6 & 8.4 & 58.8 & 54.0\tabularnewline
 & Concise &  \textcolor{green}{\cmark}  &  \textcolor{green}{\cmark}  & 27.0 & 59.4 & 30.4 & 14.7 & 55.8 & 57.6\tabularnewline
 & CodeExecutor &   \textcolor{green}{\cmark} &  \textcolor{green}{\cmark}  & 24.9 & 59.4 & 32.6 & 9.7 & 57.0 & 55.2\tabularnewline
 & NeXT & \textcolor{green}{\cmark}   &  \textcolor{green}{\cmark}  & 29.1 & 61.4 & 30.6 & 16.0 & 56.9 & 52.8\tabularnewline
 % & \ourtrace &  \textcolor{green}{\cmark}  & \textcolor{green}{\cmark}   & 27.2 & 55.4 & 32.4 & 10.5 & 56.6 & 55.9\tabularnewline
 & SemCoder(GPT4o) & \textcolor{green}{\cmark}   &  \textcolor{green}{\cmark}  & 22.2 & 59.4 & 31.4 & 10.9 & 58.6 & \underline{58.0}\tabularnewline
 & SemCoder &  \textcolor{green}{\cmark}  & \textcolor{green}{\cmark}   & \underline{29.4} & 61.9 & 33.4 & 14.7 & 59.9 & 55.4\tabularnewline
\midrule 
\multirow{9}{*}{Gemma2} & - & \textcolor{red}{\xmark} &\textcolor{red}{\xmark}  & 20.9 & \underline{63.7} & \underline{29.8} & \underline{32.8} & 49.2 & 41.5\tabularnewline
 & only NL2Code &  \textcolor{green}{\cmark} & \textcolor{red}{\xmark}  & 19.8 & 61.4 & 26.8 & 12.6 & 57.9 & 55.6\tabularnewline
 & w/o trace &  \textcolor{green}{\cmark}  &  \textcolor{red}{\xmark}  & 24.9 & 58.4 & 25.1 & 6.7 & 57.8 & 57.5\tabularnewline
 & Concise & \textcolor{green}{\cmark}   &   \textcolor{green}{\cmark} & 22.8 & 60.2 & 28.8 & 8.0 & 57.6 & 57.2\tabularnewline
 & CodeExecutor &  \textcolor{green}{\cmark}  &  \textcolor{green}{\cmark}  & 22.8 & 59.4 & 27.3 & 8.8 & 58.9 & \underline{58.2}\tabularnewline
 & NeXT &  \textcolor{green}{\cmark}  &  \textcolor{green}{\cmark}  & 26.2 & 58.1 & 26.9 & 8.8 & \underline{59.5} & 55.8\tabularnewline
 % & \ourtrace & \textcolor{green}{\cmark}   & \textcolor{green}{\cmark}   & 25.1 & 56.9 & 29.0 & 8.0 & 58.9 & 56.2\tabularnewline
 & SemCoder(GPT4o) &  \textcolor{green}{\cmark}  &  \textcolor{green}{\cmark}  & 24.1 & 62.9 & 29.5 & 8.4 & 58.9 & 56.5\tabularnewline
 & SemCoder & \textcolor{green}{\cmark}   & \textcolor{green}{\cmark}   & \underline{26.2} & 62.2 & 27.6 & 13.0 & 58.9 & 56.8\tabularnewline
\bottomrule
\end{tabular}

}
\caption{Evaluation results for full-parameter fine-tuning with semantic information on three different base models across three downstream tasks (code repair, code synthesis, and code reasoning). The “trace” setting indicates whether the LLM output includes semantic information. “only NL2Code”: fine-tuning using only code generation data without code repair data. “w/o trace”: fine-tuning with both code generation data (i.e., NL2Code) and code repair data, where the execution trace is not included in the code repair data. We report pass@1 under greedy decoding, following each benchmark’s recommended settings. BigCodeBench measured on the full set and LiveCodeBench is on the easy subset. The best scores per model are \underline{underlined} }
\label{tbl:rq1_overview}
\end{table*}

% \begin{table}[t]
% \centering
% \resizebox{0.5\textwidth}{!}
% {
% \begin{tabular}{|c|c|c|c|c|c|c|}
% \hline 
% task &  &  &  &  &  & \tabularnewline
% \hline 
% (w/ > w/o trace) / total number & 27(win\_has\_trace) & 180(total) &  &  &  & \tabularnewline
% \hline 
% The number of times each trace method achieves Top 1.  & concise: & codeexecutor & next & our & sem\_gpt4 & sem\tabularnewline
% \hline 
%  & 2 & 5 & 3 & 4 & 5 & 8\tabularnewline
% \hline 
% The number of times semcoder with GPT-4 outperforms the original semcoder. & 6 & 10(total) &  &  &  & \tabularnewline
% \hline 
% The number of times semcoder with GPT-4 y outperforms the original semcoder. & 8 & 10(total) &  &  &  & \tabularnewline
% \hline 
% semcoder with gpt4 > semcoder with gpt4 y & 4 & 10(total) &  &  &  & \tabularnewline
% \hline 
% \end{tabular}
% }
% \end{table}

\textbf{Comparison between fine-tuning with and without semantic information.} Table~\ref{tbl:rq1_overview} summarizes the performance of Code LLMs after fine-tuning. Surprisingly, the results demonstrate that fine-tuning with trace information cannot enhance the performance of Code LLMs. Specifically, for \textit{Program Repair} tasks, compared to models trained without traces~(w/o trace), only SemCoder contributes to fine-tuning but with limited improvements~(from 0.3 to 1.4). Similarly, the results of \textit{Code Synthesis} tasks show that semantic information cannot significantly enhance the code generation ability of Code LLMs. In more than half of the cases~(7 out of 9 cases), fine-tuning without trace information achieves the best results. Besides, there is also a similar phenomenon in the \textit{Reasoning} tasks.

\textit{Takeaway: }Integrating trace-based semantic information into the fine-tuning datasets cannot significantly enhance the code generation capability of Code LLMs.

\textbf{Comparison between different trace representations.} We then investigate whether there is a trace representation that is relatively better than others. Unfortunately, the results demonstrate that no single trace representation consistently outperforms others. Considering different tasks separately, \textit{SemCoder} is the best choice for program repair tasks, and \textit{SemCoder~(GPT4o)} can consistently enhance the reasoning ability of Code LLMs. \textit{Takeaway:} \textit{SemCoder} and \textit{SemCoder~(GPT4o)} are recommended representations used in fine-tuning for program repair and code reasoning tasks.

\begin{figure*}[t]
    \centering
    % \textwidth inside figure* already refers to the full‐width of both columns
    \includegraphics[width=.8\textwidth]{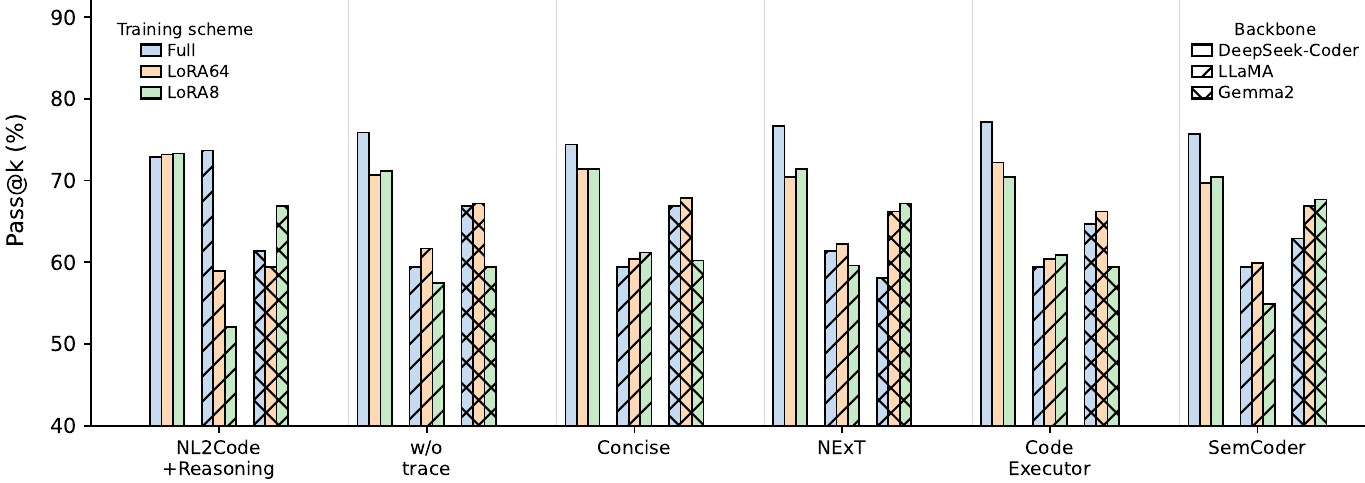}
    \caption{Fine-tuning using different training methods, i.e., Full, LoRA64, and LoRA8.}
    \label{fig:2x3images}
\end{figure*}

%%%%%%%%%%%%%%%%%
%%%%%%%%%%%%%%%%%
%%%%%%%%%%%%%%%%%
%%%%%%%%%%%%%%%%%

\subsection{Parameter-Efficient Fine-Tuning}

Parameter-efficient fine-tuning~(e.g., LoRA) is widely applied for LLMs. In this part, we explore the influence of LoRA on the fine-tuning of Code LLMs considering semantic information. 

Figure~\ref{fig:2x3images} depicts the performance of fine-tuned Code LLMs, the detailed results can be found in Appendix~\ref{more_rq1}. The results indicate that the effectiveness of parameter-efficient fine-tuning is model-dependent. Concretely, fully fine-tuning performs the best for DeepSeek model~(in 5 out of 6 cases), but LoRA enhances the model performance of LLaMA and Gemma2 in most cases (11 out of 12 cases). Similarly, LoRA8 and LoRA64 perform inconsistently across different models, and it is hard to justify which strategy is better. Furthermore, training methods highly affect the ability of code LLM trained, the performance gap between different methods can be up to 21.6 (model LLaMA). This reminds us that choosing proper training methods is crucial for Code LLMs.

% From the results, we can see that generally, LoRA64 outperforms LoRA8 and fully fine-tuning across Llama3.1-8B and Gemma2-9B models. For example, gemma2-9B + LoRA64 achieves 12.95\% overall pass@1 (vs vanilla 11.4\%), while llama3.1-8B + LoRA64 reaches up to 18.8\% (vs vanilla 17.4\%). However, model architecture is also an influence factor, gemma2-9B occasionally shows marginal gains with LoRA8 (e.g., 13.07\% pass@1 vs LoRA64’s 12.95\%), suggesting rank optimization depends on the base model. 

Considering different trace representations, the results confirm our previous conclusion that trace-based semantic information cannot significantly enhance the performance of Code LLMs through fine-tuning. However, we found that Gemma2-9B is better adapted to traces, achieving competitive results with strategies such as Semcoder (10.34\% pass@1) while maintaining repair improvements. Besides, LoRA64 without trace information is best for general code generation, while LoRA64 + repair-focused traces (e.g., Semcoder\_GPT4) maximizes repair capabilities.

\textit{Takeaway:} Parameter-efficient training methods significantly affect the performance of Code LLMs. However, the effectiveness of each method is highly model-dependent. Besides, fine-tuning without semantic information is still the best choice for preparing Code LLMs with better performance when considering these methods.

% is superior overall, but base model architecture (e.g., gemma2-9B vs llama3.1-8B) impacts rank effectiveness.  
% \textbf{Takeaway} Traces enable specialized gains (e.g., repair, hard tasks) but trade off broader performance (e.g., code generation, easy tasks).  
% \textbf{Takeaway} Task-driven tuning: Use LoRA64 + traces for niche objectives (e.g., Semcoder for hard problems) and LoRA64 alone for balanced code generation.

\subsection{Inference Test-Scaling Computation }

\begin{table*}[t]
\centering
\resizebox{1.\textwidth}{!}
{

\begin{tabular}{l|cc|ccccc|ccccc}
\toprule 
\multirow{2}{*}{} & \multirow{2}{*}{Greedy} & \multirow{2}{*}{COT} & \multicolumn{5}{c}{Sequential Scaling } & \multicolumn{5}{c}{Parallel Scaling }\tabularnewline
 &  &  & w/o trace & CodeExcutor & Concise & NExT & SemCoder & w/o trace & CodeExcutor & Concise & NExT & SemCoder\tabularnewline
\midrule 
% \textit{Close source Model}\\
GPT-4o-mini & 73.08 & 73.08 & 98.46 & 98.46 & 99.23 & 99.23 & 99.23 & 88.46 & 80.77 & 80.77 & 84.62 & 80.77\tabularnewline
deepseek-chat(V3) & 84.62 & 100.00 & 100.00 & 100.00 & 100.00 & 100.00 & 100.00 & 96.15 &96.15  &96.15  & 96.15 & 92.3 \tabularnewline
\midrule 
\textit{Reasoning Compatible Model}\\
AIDC-AI/Marco-o1 & 53.85 & 50.00 & 76.92 & 69.23 & 76.92 & 73.08 & 73.08 & 61.54 & 53.85 & 69.23 & 61.54 & 57.69\tabularnewline
microsoft/phi-4 & 53.85 & 73.08 & 100.00 & 96.15 & 100.00 & 91.54 & 100.00 & 80.77 & 76.92 & 84.62 & 80.77 & 84.62\tabularnewline
\midrule 
\textit{Instruction of Foundation Model}\\
Llama-3.1-8B-Inst & 34.62 & 34.62 & 67.69 & 66.92 & 74.62 & 74.62 & 65.38 & 46.15 & 42.31 & 57.69 & 57.69 & 57.69\tabularnewline
deepseek-coder-6.7b-Inst & 42.31 & 46.15 & 68.46 & 61.54 & 69.23 & 76.15 & 69.23 & 53.85 & 50.00 & 57.69 & 61.54 & 50.00\tabularnewline
Qwen2.5-Coder-7b-Inst & 61.54 & 34.62 & 83.85 & 87.69 & 80.77 & 90.77 & 86.92 & 53.85 & 61.54 & 65.38 & 50.00 & 53.85\tabularnewline
\bottomrule 
\end{tabular}

}

\caption{Pass@\!1 accuracy on the \textbf{LiveCodeBench} (\textit{easy}) private test set under equal compute budgets. Values are the percentage of prompts whose \textit{final} completion passes \textit{all} private test cases. \textbf{Greedy}: one-shot, highest-probability decode. \textbf{CoT}: answer plus natural-language rationale. \textbf{Sequential Scaling}: 8 parallel samples (T = 0.7) followed by $R_{max}$=4 self-debugging rounds on public tests, selecting the best candidate. \textbf{Parallel Scaling}: 16 candidates ranked by votes from an LLM-as-a-Judge on execution result along with its trace representation.  The “\textit{w/o trace}” variants rely only on the initial execution output, whereas trace-based variants leverage execution traces representations during self-debugging or voting. Sequential Revision benefits most from trace‐aware signals. Double underlining marks the overall best LiveCodeBench private-set score.}

\label{tbl_scaling_inference_new}
\end{table*}

Table~\ref{tbl_scaling_inference_new} summarizes the results of test-time scaling. It is clear that, compared to open-source LLMs, closed-source LLMs perform significantly better at test time.

\textbf{Impact of test-scaling strategies.} First, the results demonstrated that test-scaling consistently enhances the code generation ability of Code LLM. Concretely, in 65 out of 70 cases, test scaling strategies achieved higher Pass@1 scores than Greedy and COT.  \textbf{Impact of semantic representation.}  The results suggest that, similar to SFT, the usefulness of semantic information is blurred. In more than half cases~(36 out of 56 cases), integrating semantic information into the input prompt cannot help Code LLM to produce correct code compared to without adding semantic information. However, one semantic representation~(\textit{Concise}) stands out, which achieved Pass@1 no worse than \textit{w/o trace} in 11 out of 14 cases.

\textit{Takeaway:} Similar to fine-tuning, most of the trace-based semantic representations cannot enhance the performance of Code LLMs at test time except for \textit{Concise}.

% Different from fine-tuning, trace-based semantic information significantly enhances the performance of Code LLMs at test time. Sequential Revisions and BeamSearch are two optimal search strategies for test-time scaling.

\subsection{Hyperparameter Study~\label{sec:hyper}}

We further explore the impact of hyperparameters on Sequence scaling, which significantly boosts Code LLMs at test time. 

\begin{figure}[t]
    \small
	\centering
	\includegraphics[width=.75\columnwidth]{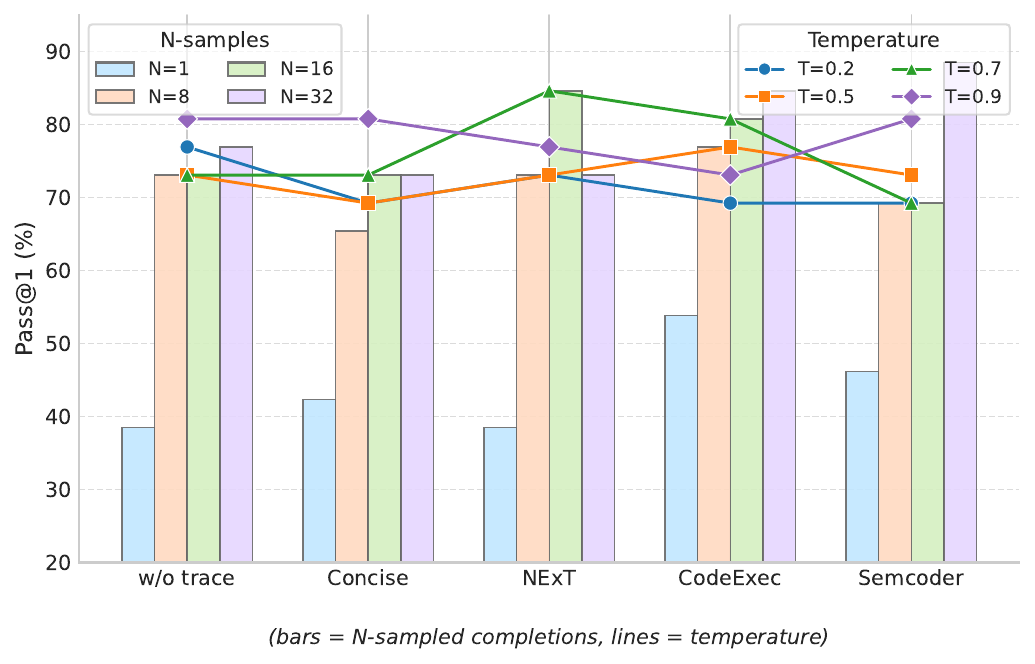}
	\caption{ Pass\texttt{@}1 of Qwen-7B on the LCB "easy" split as a function of sampled completions N (bars) and decoding temperature T (lines) across five trace formats. Two trends emerge: (i) increasing N yields substantial gains across all formats; (ii) higher temperatures (T > 0.7) generally outperform lower ones. NExT and CodeExecutor achieve the best results (88.5\% at N=32, T=0.9), followed by Semcoder, while the baseline w/o trace consistently underperforms.}
	\label{fig:effect_n}
    \vspace{-2mm}
\end{figure}

We first investigate the impact of model temperature. Figure~\ref{fig:effect_n} illustrates the results, where Pass@1 scores fluctuate under different temperatures. One conclusion we can draw is that a small temperature~(T=0.2) negatively affects the performance of Code LLMs, higher temperature performs relatively better. 

Sequence scaling has two parameters, the iteration number~(\textit{rounds}) and the generated samples~(\textit{samples}) during each iteration. The results in Figure~\ref{fig:effect_n} show that the more samples generated, the higher Pass@1 scores achieved by Code LLMs. However, there is a trade-off between the sample numbers and the performance of Code LLMs. \textit{samples}=8 is the default setting in our framework. The results of \textit{rounds} study can be found in Table~\ref{tab:errtype}, where more rounds lead to better code generation capability of Code LLMs.

%% file: sections/related_works.tex
\paragraph{Chain-of-Thought (CoT) and Tool-Integrated Reasoning (TIR).}  

Beyond execution traces, recent advances emphasize explicit reasoning steps and tool usage. \textit{Chain-of-Thought (CoT)}~\cite{wei2022chain} enables LLMs to decompose complex problems into intermediate reasoning steps, improving accuracy in tasks like mathematical problem-solving. However, some tasks require computational precision beyond language reasoning. \textit{Tool-Integrated Reasoning (TIR)}~\cite{gou2023tora} addresses this by integrating LLMs with external tools (e.g., Python interpreters) for specialized computations, excelling in symbolic computation and high-complexity algorithms. These approaches highlight the trend of augmenting LLMs with runtime observations or external tools, which our work builds upon by systematically evaluating their impact on reasoning and code generation. 
% Different from previous works, we are the first to study the usefulness of TIR on code generation tasks. 

\paragraph{LLMs for Software Engineering and Execution Behavior.}  
Code execution behavior encompasses runtime information (e.g., program state, execution paths) and pre-/post-execution details. Recent studies leverage these behaviors to enhance LLM performance. For instance, ~\cite{chen2023teaching} introduced Self-Debugging, where LLMs generate explanations to guide debugging; ~\cite{ni2024next} proposed NExT, representing execution behaviors as inline comments for fine-tuning; and ~\cite{ding2024semcoder} described runtime behaviors in natural language for LLM training. While prior works typically use a single representation, our study explores multiple execution-based representations and their impact on code generation and reasoning tasks.

LLMs are widely applied in software engineering, including vulnerability detection~\cite{shestov2024finetuning}, bug repair~\cite{xia2023keep,Wang2024RAT,Wang2025DBL}, and code generation~\cite{hong2023metagpt,wu2023autogen,tang2024collaborative}. Evaluation frameworks (e.g., EvalPlus~\cite{liu2023is}) and datasets (e.g., ClassEval~\cite{du2024evaluating}, SWE-bench~\cite{jimenez2023swe}) have been developed to benchmark these capabilities. While prior efforts focus on task-specific performance, our work investigates how execution-centric signals, inspired by human debugging, enhance LLMs' proficiency in code generation and reasoning.

% Different from the above works, we thoroughly evaluate the usefulness of trace-based code semantic information on both SFT and inference of Code LLMs.

\paragraph{Scaling Up Inference-Time Computing}  
Recent advances in inference-time computing have improved the verification of mathematical reasoning in LLMs. ~\cite{cobbe2021training} introduced token-level reward models to score individual steps, while ~\cite{xiao2024densing} refined these with process reward models (PRM) for granular feedback. ~\cite{snell2024scaling} demonstrated that scaling inference-time computing is more cost-effective than retraining models. Building on these, we sample multiple solutions from LLM reasoners and explore verifier training approaches. Our framework, adapted from~\cite{sky_t1_2025}, systematically evaluates how the semantic information impacts the post-training, one-time inference, and scaling inference test-time with runtime behavior can achieve strong performance in code-related tasks.

% \begin{table}[ht]
% \centering
% \begin{tabular}{lcccccccc}
% \toprule
% \textbf{Test-case size} &
% \multicolumn{2}{c}{\textbf{exe}} &
% \multicolumn{2}{c}{\textbf{concise}} &
% \multicolumn{2}{c}{\textbf{next}} &
% \multicolumn{2}{c}{\textbf{our}}\\
% \cmidrule(lr){2-3}\cmidrule(lr){4-5}\cmidrule(lr){6-7}\cmidrule(lr){8-9}
%  & col-1 & col-2 & col-3 & col-4 & col-5 & col-6 & col-7 & col-8\\
% \midrule
% size=1 &
% $32.30$ & $28.20$ & $32.90$ & $28.80$ &
% $31.80$ & $27.90$ & $32.90$ & $29.00$ \\[2pt]

% size=2 &
% $32.90_{\scriptscriptstyle +0.6}$ &
% $29.60_{\scriptscriptstyle +1.4}$ &
% $31.80_{\scriptscriptstyle -1.1}$ &
% $28.80_{\scriptscriptstyle 0.0}$ &
% $31.80_{\scriptscriptstyle 0.0}$ &
% $28.80_{\scriptscriptstyle +0.9}$ &
% $32.60_{\scriptscriptstyle -0.3}$ &
% $29.60_{\scriptscriptstyle +0.6}$ \\[2pt]

% size=3 &
% $32.60_{\scriptscriptstyle +0.3}$ &
% $28.80_{\scriptscriptstyle +0.6}$ &
% $29.80_{\scriptscriptstyle -3.1}$ &
% $27.80_{\scriptscriptstyle -1.0}$ &
% $23.90_{\scriptscriptstyle -7.9}$ &
% $23.90_{\scriptscriptstyle -4.0}$ &
% $33.70_{\scriptscriptstyle +0.8}$ &
% $30.40_{\scriptscriptstyle +1.4}$ \\
% \bottomrule
% \end{tabular}
% \caption{Performance () with percentage-point gains (subscripts show change from the \textit{size = 1} baseline).}
% \end{table}

% This study empirically evaluates the contribution of each component, showing that integrating these signals significantly improves accuracy and interpretability across software engineering and mathematical reasoning tasks.

%% file: sections/p5_conclusion.tex
This paper introduces a generic framework for generating trace-based code semantic information. Based on this framework, we systematically evaluate the usefulness of trace-based code semantic information for fine-tuning and inference of Code LLMs. The results highlight that existing code semantic information does not enhance fine-tuning and test-time scaling of Code LLMs. 
% This work can serve as the new baseline for the study of leveraging semantic information to enhance Code LLMs.

This work can serve as the new baseline for the study of leveraging semantic information to enhance Code LLMs. This opens up several avenues for future research. First, it is essential to design new forms of semantic representations that are more aligned with how Code LLMs process and understand code, potentially incorporating higher-level abstractions or contextual cues. Second, future work should explore more effective strategies for integrating semantic information into model training and inference pipelines -- such as architectural modifications, specialized pretraining objectives, or more adaptive prompting techniques.

%% file: sections/limitation.tex
\section*{Limitations}

\textbf{Limited Programming language supported.} Currently, our framework only supports Python, other programming languages, such as Java and C++, are not supported. Even though, we believe our findings can provide insights to developers who plan to enhance their Code LLMs via semantic information interaction. Besides, we plan to actively maintain our framework to cover more programming languages. 

\textbf{Limited LLM size.} Due to constraints on computational resources, we only conduct experiments on LLMs with around 7B size. Experiments with larger LLMs could be our future work.

\section*{Ethical Considerations}

\paragraph{Research purpose and societal impact.}
This project seeks to deepen scientific understanding of how dynamic program semantics influence Code-LLM reasoning, with the ultimate goal of producing safer, more reliable coding assistants. All artefacts—datasets, code, and models—are released solely for research and evaluation; they are \textbf{not} intended for autonomous deployment in production settings.

\paragraph{Provenance, licensing, and consent.}
Source code used for fine-tuning and evaluation is drawn exclusively from repositories under OSI-approved permissive licences (e.g., MIT, Apache-2.0).  No proprietary material is incorporated without explicit permission.

\paragraph{Privacy preservation in execution traces.}
Because runtime logs can inadvertently expose credentials or personally identifiable information (PII), every trace passes a three-stage sanitation pipeline: (i) static pattern-based redaction of common secret/PII formats; (ii) dynamic taint tracking that masks values originating from environment variables, network sockets, or file I/O; and (iii) manual review of a random \(2\%\) sample per release. Traces failing any check are discarded.

% \paragraph{Dual-tier data release strategy.}
% To balance transparency with security, we publish two versions of each trace set:\\
% \textbf{Public}: summarised control-flow hashes and bounded value ranges—sufficient for benchmarking but insufficient to reconstruct full program logic.\\
% \textbf{Restricted}: full line-level traces (calls, locals, identifiers) available only to vetted academic partners who sign a security addendum pledging safe handling and non-redistribution.

\paragraph{Safeguards against malicious use.}
Enhanced semantic reasoning could facilitate the generation of vulnerable or harmful code. We therefore (i) withhold weights fine-tuned on explicitly security-sensitive benchmarks, (ii) deploy server-side filters that block outputs matching exploit-related patterns (shell execution, SQL/command injection, path traversal), and (iii) document residual unsafe generations in an appendix to encourage community development of stronger guards.

\paragraph{Bias and inclusivity in dataset design.}
Although our empirical study concentrates on mainstream languages, we provide \emph{PySnooper~\cite{rachum2019pysnooper}} for Python, and Solidity to spur broader, community-driven extension. We encourage downstream researchers to audit biases that may arise when applying our framework to new ecosystems or developer populations.

% \paragraph{Energy use disclosure.}
% Trace instrumentation and test-time scaling add reasonable compute per model–dataset pair. We offset these emissions through Gold-Standard renewable-energy credits and release all scripts so others can reproduce results with fewer redundant runs.

\paragraph{Responsive governance.}
We publish our redaction and inspection pipeline under an open-source licence, provide a dedicated security-contact email for vulnerability disclosures, and commit to removing or revising any resource within 30 days of a substantiated harm report.

\bigskip
\noindent
In summary, we have instituted licensing checks, consent agreements, privacy filters, controlled release mechanisms, and transparent governance to ensure that the benefits of semantics-aware Code-LLMs are realised responsibly while potential harms are proactively mitigated.

\clearpage

%% file: sections/p6_appendix_1.tex
% \textbf{Models.}\wj{Qiang: please review} We perform fine-tuning on three base LLMs: deepseek-6.7b-base, Llama3.1-8B, and gemma2-9b, which we selected for their non-trivial yet unsaturated performance across code and reasoning tasks~\cite{ding2024semcoder,snell2024scaling}. For inference, we evaluate on phi-4~\cite{abdin2024phi} and Marco-O1~\cite{zhao2024marcoo1openreasoningmodels} (two reasoning-oriented models), as well as CodeLlama-7B-Instruct, deepseek-6.7b-instruct and Llama3.1-8B-instruct (three instruction-tuned foundation models) and two Close-source general model( GPT-4o and DeepseekV3, i.e Deepseek-Chat). Together, these models span a wide range of capabilities in both reasoning and instruction settings, making them representative of many contemporary LLMs and an excellent test bed for examining the benefits of semantic refinement under a fixed compute budget.
\section{ Experiment }\label{exp_config}

\subsection{Experiment Detail. }

\paragraph{Finetune} We conducted SFT experiments on three different base models using two distinct configurations. First, we perform full-parameter tuning with DeepSpeed ZeRO-3 using learning rates of 2.0e-5; training batch sizes of 8; bfloat16 precision; a maximum sequence length of 2048; and two epochs. Next, we explore LoRA-based tuning with various LoRA ranks, disabling DeepSpeed while maintaining the same bfloat16 precision, maximum sequence length, and epoch count.

\paragraph{Scaling inference.}

\textit{Sequential Revisions} uses an external tool, a Python interpreter, to verify each predicted solution. As introduced in Section~\ref{sec:dataconstruct}, the feedback from tools is appended to the prompt for the next generation iteration. This iterative process continues until a successful solution emerges or the compute budget of $R_{\max}=8$ rounds is reached.

\textit{Parallel Scaling} generates $N$ diverse candidate solutions in parallel to increase the likelihood of finding a correct solution. We synthesize test inputs, execute all candidates, and collect their execution outputs and trace representations as scoring prompts. An LLM-as-a-judge then ranks solutions based on these scoring prompts and pre-trained knowledge. This Parallel Scaling complements Sequential Revisions to maximize code generation capabilities without additional training. Specifically, we sample $N=16$ answers independently from the specified LLM and then select the best answer according to the reward model's final judgment. We use Phi-4 as the reward LLM to score each candidate with a reward function (e.g., static analysis, unit tests, or a learned model, similar to Best-of-N sampling), and retain only the top $\frac{N}{M}$ candidates. We score these again, prune to the top-ranked solutions, and repeat iteratively until all buggy lines are addressed. The result is up to $N$ complete repaired-code solutions, from which we select the best via a final evaluation, always choosing the highest score as judged by the reward model.

\textit{Greedy} sampling generates only 1 answer independently from the specified LLM by setting the temperature to 0.

\textbf{Implementation and Environment.} We implement all Code LLMs based on Hugging Face APIs, the implementation of the fine-tuning process is modified from the official project of~\cite{hu2021lora}. We use OpenAI's official APIs to access gpt-3.5-turbo-0125/gpt-4o-mini-2024-07-18/gpt-4o-2024-05-13 models. Finetune experiments have been conducted on eight NVIDIA A100 GPUs using the Distributed Data Parallel (DDP) module. Inference jobs utilize the vLLM~\cite{vllm}, which is a unified library for LLM serving and inference.

%% file: sections/p6_appendix_2.tex
\section{ Dataset }\label{exp_dt}

\subsection{Finetune Dataset Quality Filtering and Decontamination}
We follow the ~\cite{wei2024selfcodealign} to conduct Decontamination and Refinement.

\paragraph{Removing Benchmark Data}
To ensure the integrity of our evaluation process, we rigorously decontaminated the dataset by removing any functions that resembled prompts or solutions from the benchmarks used for evaluation. This step is critical to prevent data leakage and ensure a fair assessment of our method. Specifically, we checked for the presence of substrings from benchmark prompts or solutions within the dataset. Any function containing such substrings was excluded. This process guarantees that the dataset remains unbiased and does not inadvertently include examples that could skew evaluation results.

\paragraph{Docstring Quality Filtering}
We observed that many Python functions, while containing docstrings, often had poor or misleading documentation. To address this, we employed StarCoder2-15B, a state-of-the-art language model, to perform binary classification on the docstrings. The model was tasked with identifying functions with low-quality or misleading documentation. Functions flagged as having poor docstrings were removed from the dataset. This step ensures that the retained functions are not only functional but also well-documented, enhancing their usability for downstream tasks such as code understanding and generation.

In sum up, the decontamination and refinement process, particularly the removal of benchmark-related data and the filtering of low-quality docstrings, plays a pivotal role in ensuring the quality and reliability of our dataset. By meticulously removing functions that could compromise evaluation fairness and those with inadequate documentation, we have created a robust dataset of 248,934 high-quality Python functions. This dataset is well-suited for a wide range of applications, including code generation, evaluation, and analysis, while maintaining a high standard of integrity and usability.

\subsection{Evaluation dataset}

\textbf{Evaluation on Code Generation and Reasoning Tasks}
In this study, we fine-tune Code LLMs using the refinement dataset described in Section~\ref{sec:refinementdataset}. For the experiment, we deploy the fine-tuned models in two code generation tasks: program repair and code synthesis. We further fine-tune the Code LLMs specifically for reasoning tasks to assess whether their reasoning capabilities are enhanced. We evaluate the performance of the fine-tuned models using open-source test datasets.

Note that to mitigate potential data leakage risks, we adhere to established methods as outlined in~\cite{muennighoff2023octopack} by conducting a thorough decontamination process. This ensures that there is no overlap between our fine-tuning dataset and the evaluation datasets utilized.

\begin{itemize}

\item \textit{Program Repair.} 
To evaluate program repair capabilities, we construct datasets of buggy code using benchmarks from HumanEval, MBPP, and CRUXEval, maintaining a consistent sample size of 164, which aligns with HumanEval. For HumanEval, we directly use buggy code from the existing dataset HumanEvalPack~\cite{humanevalpack}. For the other two datasets, which contain larger sample sizes, we randomly select 164 samples from each. Following the methodology described in~\cite{ni2024next}, we employ GPT-4, GPT-3.5-Turbo, and CodeLlama-34B to generate solutions for each problem. From these, we select one incorrect solution per problem based on test case validation. This process results in collections of 164 buggy code samples for each dataset, denoted as Human-R, MBPP-R, and CRUXEval-R.  For the repair evaluation, the prompts provided include the buggy code, the corresponding failed test case, and the execution traces of that test case \footnote{Due to space constraints, we have made all prompt templates, including those used for fine-tuning and evaluation, available on our website.}.

\item \textit{Code Synthesis.} 
We evaluate the code synthesis capabilities of the tuned Code LLMs using two widely-used datasets, HumanEval and MBPP, both provided by EvalPlus~\cite{liu2023is}. To ensure consistency in our assessment, we employ the same prompts and pre-processing methods as outlined in~\cite{liu2023is}. Additionally, we differentiate between two sets of test cases from EvalPlus, referred to as \textit{base} and \textit{plus}, in our evaluations.

\item \textit{Code Reasoning.} 
We follow the existing works~\cite{chen2024evaluating,gu2024cruxeval,Wang2024AES} to conduct the reasoning tasks, i.e., input prediction, output prediction, state prediction, and coverage prediction. Input/output prediction indicates fulfilling the corresponding input or output, given a block of code and a partially completed assertion statement as a prompt. State prediction refers to predicting what the next line statement will be after an intermediate statement is executed. Coverage prediction means that after randomly picking a line of code,  we ask the LLM to predict whether it will be executed for a given specific test case. For input and output prediction, we directly use the existing datasets~\cite{gu2024cruxeval}, named as CRUXEval-I and CRUXEval-O. For the evaluation, we follow the same prompts from~\cite{chen2024evaluating,gu2024cruxeval}.

\end{itemize}

\subsection{Fine-tune dataset(Refinement Dataset)} \label{sec:refinementdataset}

We found that there are no datasets that fully support our study, i.e., the repair-based training mode and all types of execution representation that we collected. Hence, we construct a new dataset that covers buggy code, its corresponding patch, test cases, and other semantic information such as execution traces.  
% To study the usefulness of our framework and the effectiveness of different execution traces, and more importantly, to further facilitate this research direction, we construct the first dataset that covers buggy code, its corresponding patch, and other semantic information such as execution traces.  

Our dataset is constructed using APPs~\cite{hendrycksapps2021}, a dataset provided by codeparrot for the generation of code at the competition level. This dataset includes essential elements such as basic buggy code, correct code, test cases, and the human refinement trajectories from buggy to correct versions.

The steps to construct the dataset are as follows:
\begin{enumerate}
\item \textit{Buggy and Patch Pair Collection}: Each problem in apps includes multiple solutions, both correct and incorrect, provided by various users. A key challenge in extracting (buggy code, patch code) pairs is the difficulty in matching incorrect code with corresponding correct solutions due to anonymized author information for privacy. To overcome this, we employ a similarity-based matching approach, as the buggy code and its refined version from the same author typically exhibit significant similarities. Specifically, we employ UniXcoder~\cite{guo2022unixcoder} to extract embeddings of both correct and incorrect programs and calculate their embedding similarity using Cosine similarity measurement. We include pairs with a similarity score above 0.8 in our dataset as  ($B_X$, $P_Y$). 

\item \textit{Test Case Extraction}: After collecting the code pairs, we execute both the buggy and patched code using their accompanying test cases from apps. We retain test cases that fail with the buggy code but pass with the patched code, designating them as failing test cases, which are then incorporated into $R_X$.

\item \textit{Execution Trace Extraction}: We execute both the buggy and patched codes under the failing test cases and use Trace-Tracker, a Python debugging tool, to gather runtime information, including trace coverage and states. We develop converters to translate this runtime information into various trace representations (detailed in Section 3.1). These traces are added to $R_X$ and $R_Y$, respectively.

\item \textit{Code Reasoning Related Data Extraction}: Leveraging the execution traces, we enrich our dataset with features specifically designed to evaluate various aspects of code reasoning: input prediction, output prediction, state prediction, and coverage prediction. For input and output predictions, we adhere to the methods outlined in \cite{gu2024cruxeval}, inserting assert statements to validate the inputs and outputs effectively. 

\item \textit{Program Description Extraction}: Following previous work~\cite{codecontest}, we utilize each contest problem's brief description as a base. We then employ GPT-4o to enrich these descriptions by generating implementation constraints and incorporating test cases.

\end{enumerate}

%% file: sections/p6_appendix_3.tex
% \clearpage
\section {Experiment results of Full and PEFT }\label{more_rq1}

\begin{table*}[t]
\small
\centering
\resizebox{0.95\textwidth}{!}
{

\begin{tabular}{cc|cc|cc|cc|cc|cc}
\multirow{2}{*}{} &  & \multicolumn{2}{c|}{repair} & \multicolumn{2}{c|}{NL2Code} & \multicolumn{2}{c|}{reasoning} & \multicolumn{2}{c|}{bigcodebench} & \multicolumn{2}{c}{Livecodebench}\tabularnewline
 &  & HE-R/(+) & MBPP-R/(+) & HE/(+) & MBPP/(+) & in\_predict & out\_predict & full & hard & easy pass@1 & overall pass@1\tabularnewline
\midrule 
DeepSeek-Coder &  & 26.2(22.0) & 17.7(15.6) & 49.4(43.9) & 71.9(57.6) & 40 & 40.4 & 41.50 & 12.20 & 40.80 & 17.40\tabularnewline
\midrule 
\multirow{3}{*}{only nl2code} & Full & 59.1(52.4) & 25.4(22.0) & 64.6(56.1) & 72.9(60.9) & 60.1 & 55.4 & 43.70 & 16.90 & 12.60 & 4.60\tabularnewline
 & LoRA64 & 48.2(42.1) & 25.4(22.0) & 57.3(48.8) & 73.2(59.1) & 44.8 & 47.9 & 44.70 & 14.90 & 44.10 & 18.50\tabularnewline
 & LoRA8 & 44.5(37.2) & 22.2(19.3) & 52.4(44.5) & 73.3(58.1) & 42.9 & 47.2 & 44.50 & 12.20 & 41.20 & 17.40\tabularnewline
\midrule 
\multirow{3}{*}{concise} & Full & 45.7(39.0) & 39.2(33.1) & 61.6(54.3) & 74.4(61.9) & 61.6 & 55 & 44.30 & 17.60 & 29.40 & 12.60\tabularnewline
 & LoRA64 & 46.3(41.5) & 33.9(28.8) & 53.7(47.6) & 71.4(59.1) & 48.1 & 48.8 & 42.50 & 12.20 & 40.30 & 16.70\tabularnewline
 & LoRA8 & 42.7(38.4) & 35.7(30.4) & 54.3(48.2) & 71.4(59.1) & 54.1 & 48 & 45.60 & 16.20 & 38.70 & 16.30\tabularnewline
\midrule 
\multirow{3}{*}{CodeExecutor} & Full & 38.4(34.8) & 38.4(33.6) & 60.4(53.7) & 77.2(62.9) & 60.4 & 56.1 & 44.60 & 20.30 & 31.50 & 13.60\tabularnewline
 & LoRA64 & 45.1(40.9) & 36.0(31.2) & 54.3(48.2) & 72.2(60.9) & 52.5 & 48.8 & 45.00 & 15.50 & 38.20 & 17.10\tabularnewline
 & LoRA8 & 49.4(43.3) & 33.9(29.1) & 54.9(47.0) & 70.4(57.6) & 46.2 & 47.8 & 45.20 & 16.90 & 39.90 & 17.10\tabularnewline
\midrule 
\multirow{3}{*}{w/o trace} & Full & 43.9(39.0) & 39.2(34.1) & 58.5(51.8) & 75.9(61.9) & 61.9 & 56.6 & 45.40 & 16.20 & 35.70 & 15.60\tabularnewline
 & LoRA64 & 49.4(45.7) & 35.2(29.4) & 53.7(46.3) & 70.7(58.6) & 54.2 & 49.1 & 46.40 & 18.90 & 43.70 & 18.80\tabularnewline
 & LoRA8 & 47.0(40.9) & 30.4(24.9) & 53.0(44.5) & 71.2(58.6) & 49.4 & 48.4 & 46.10 & 14.90 & 41.60 & 17.70\tabularnewline
\midrule 
\multirow{3}{*}{NeXT} & Full & 39.6(36.6) & 37.6(31.7) & 61.6(54.9) & 76.7(62.4) & 61.3 & 54.2 & 44.00 & 16.90 & 36.10 & 14.90\tabularnewline
 & LoRA64 & 47.6(42.1) & 34.7(29.6) & 54.9(48.8) & 70.4(58.4) & 47.4 & 49.6 & 45.40 & 16.20 & 42.40 & 17.10\tabularnewline
 & LoRA8 & 47.0(41.5) & 34.4(28.8) & 55.5(47.6) & 71.4(57.6) & 47.4 & 47.6 & 45.00 & 17.60 & 40.30 & 17.00\tabularnewline
\midrule 
\multirow{3}{*}{concise2} & Full & 43.3(37.2) & 38.9(33.1) & 59.1(51.8) & 76.7(63.2) & 60 & 55.6 & 43.90 & 18.90 & 35.70 & 15.30\tabularnewline
 & LoRA64 & 48.8(43.9) & 35.7(30.4) & 52.4(46.3) & 70.2(59.1) & 52.8 & 49.5 & 45.40 & 16.90 & 38.20 & 16.30\tabularnewline
 & LoRA8 & 48.2(42.7) & 32.8(28.6) & 53(45.1) & 70.9(57.6) & 46.4 & 49 & 43.60 & 13.50 & 41.20 & 17.50\tabularnewline
\midrule 
\multirow{3}{*}{SemcoderGPT4o} & Full & 53.7(47.6) & 37.0(32.0) & 59.1(52.4) & 75.7(63.4) & 62 & 58.1 & 45.40 & 18.20 & 31.50 & 13.70\tabularnewline
 & LoRA64 & 51.8(46.3) & 34.9(29.9) & 55.5(48.8) & 69.7(58.9) & 53.1 & 50.7 & 46.10 & 18.20 & 39.10 & 17.10\tabularnewline
 & LoRA8 & 51.2(44.5) & 30.4(26.5) & 52.4(45.1) & 70.4(56.9) & 50.5 & 49.5 & 47.20 & 18.90 & 42.00 & 17.40\tabularnewline
\midrule 
\multirow{3}{*}{SemcoderGPT4o\_y} & Full & 47.0(40.9) & 38.9(32.0) & 61.6(55.5) & 74.7(62.9) & 61.3 & 57.4 & 45.50 & 18.90 & 42.00 & 16.70\tabularnewline
 & LoRA64 & 53.0(47.6) & 32.3(28.6) & 54.3(48.8) & 72.9(60.4) & 52.6 & 49.2 & 44.70 & 16.90 & 44.50 & 18.20\tabularnewline
 & LoRA8 & 53.7(45.7) & 31.7(27.5) & 52.4(45.7) & 72.7(58.4) & 46.1 & 48.9 & 46.10 & 17.60 & 39.90 & 17.10\tabularnewline
\midrule 
\multirow{3}{*}{Semcoder} & Full & 45.7(39.6) & 40.5(35.2) & 58.5(51.8) & 76.4(63.2) & 59.5 & 55.4 & 45.70 & 20.90 & 29.00 & 12.80\tabularnewline
 & LoRA64 & 48.2(43.3) & 34.7(30.2) & 53.0(46.3) & 70.2(58.9) & 52.4 & 50.5 & 45.90 & 17.60 & 42.90 & 18.10\tabularnewline
 & LoRA8 & 48.8(43.3) & 32(27.5) & 53.7(45.7) & 70.7(57.1) & 47 & 49.8 & 46.10 & 16.20 & 41.20 & 17.50\tabularnewline
\midrule 
\multirow{3}{*}{Semcoder\_y} & Full & 46.3(39.6) & 37.6(33.1) & 60.4(54.3) & 75.4(61.9) & 59 & 57.8 & 44.10 & 16.90 & 39.10 & 16.40\tabularnewline
 & LoRA64 & 48.2(42.7) & 32.8(29.1) & 56.1(49.4) & 72.9(60.2) & 54.1 & 49.8 & 44.80 & 19.60 & 40.80 & 17.10\tabularnewline
 & LoRA8 & 51.2(44.5) & 31.7(27.2) & 53.0(45.7) & 72.2(58.9) & 49.6 & 47.6 & 45.20 & 15.50 & 41.60 & 17.50\tabularnewline
\end{tabular}

}

\caption{A extend version of Table in DeepSeek-Coder(deepseek-6.7B-base) after finetuning with semantic information. The $\_y$ token indicates that the rationale $r$ should be generated, rather than fed into the prompt. }
\label{appindex_dk_lora}
\end{table*}

\begin{table*}[t]
\small
\centering
\resizebox{0.95\textwidth}{!}
{

% Preview source code for paragraph 0

\begin{tabular}{cc|cc|cc|cc|cc|cc}
\multirow{2}{*}{} &  & \multicolumn{2}{c|}{repair} & \multicolumn{2}{c|}{NL2Code} & \multicolumn{2}{c|}{reasoning} & \multicolumn{2}{c|}{bigcodebench} & \multicolumn{2}{c}{Livecodebench}\tabularnewline
 &  & HE-R/(+) & MBPP-R/(+) & HE/(+) & MBPP/(+) & in\_predict & out\_predict & full & hard & easy pass@1 & overall pass@1\tabularnewline
\midrule 
LLaMA &  & 28.0(26.2) & 20.1(18.0) & 38.4(32.3) & 58.6(49.1) & 42.6 & 36.2 & 31.40 & 6.08 & 27.30 & 9.50\tabularnewline
\midrule 
\multirow{3}{*}{only nl2code} & Full & 58.5(52.4) & 24.9(22.2) & 65.9(56.7) & 73.7(61.2) & 60.1 & 55.9 & 44.10 & 18.20 & 18.10 & 7.40\tabularnewline
 & LoRA64 & 47.6(42.1) & 25.7(21.7) & 44.5(40.9) & 58.9(46.4) & 54.4 & 52.9 & 35.10 & 10.10 & 29.80 & 11.10\tabularnewline
 & LoRA8 & 43.9(38.4) & 23.8(20.1) & 39.6(33.5) & 52.1(42.9) & 52.8 & 51.9 & 33.70 & 9.50 & 34.50 & 11.90\tabularnewline
\midrule 
\multirow{3}{*}{concise} & Full & 30.5(29.3) & 27.0(23.8) & 47.0(43.3) & 59.4(48.9) & 55.8 & 57.6 & 30.40 & 8.80 & 14.70 & 5.20\tabularnewline
 & LoRA64 & 34.8(31.1) & 33.6(29.9) & 40.2(35.4) & 60.4(51.4) & 54.4 & 51.7 & 35.20 & 10.10 & 21.80 & 8.00\tabularnewline
 & LoRA8 & 39.6(32.3) & 31.0(27.2) & 36.6(32.3) & 61.2(49.9) & 53.2 & 51.9 & 33.90 & 7.40 & 32.80 & 11.60\tabularnewline
\midrule 
\multirow{3}{*}{CodeExecutor} & Full & 29.9(29.3) & 24.9(22.2) & 53.7(50.0) & 59.4(47.9) & 57 & 55.2 & 32.60 & 9.50 & 9.70 & 3.60\tabularnewline
 & LoRA64 & 31.1(28.0) & 32.8(28.8) & 39.6(34.8) & 58.9(49.6) & 53.9 & 51.2 & 33.90 & 6.80 & 18.90 & 6.70\tabularnewline
 & LoRA8 & 37.8(30.5) & 31.0(27.0) & 36.0(32.9) & 60.9(49.9) & 53.4 & 52.2 & 33.30 & 7.40 & 34.00 & 11.60\tabularnewline
\midrule 
\multirow{3}{*}{w/o trace} & Full & 28.0(24.4) & 29.1(26.2) & 52.4(49.4) & 59.1(47.1) & 58.8 & 54 & 31.60 & 8.10 & 8.40 & 3.10\tabularnewline
 & LoRA64 & 31.7(29.3) & 32.0(27.5) & 42.7(40.9) & 61.7(52.6) & 53.5 & 50.5 & 34.70 & 10.80 & 22.70 & 8.00\tabularnewline
 & LoRA8 & 47.0(40.9) & 29.1(25.7) & 36.6(32.9) & 57.5(47.1) & 50.2 & 49.8 & 32.50 & 9.50 & 31.50 & 11.20\tabularnewline
\midrule 
\multirow{3}{*}{NeXT} & Full & 28.7(26.2) & 29.1(25.4) & 49.4(46.3) & 61.4(49.1) & 56.9 & 52.8 & 30.60 & 6.80 & 16.00 & 5.60\tabularnewline
 & LoRA64 & 28.7(25.0) & 32.5(28.8) & 43.9(39.6) & 62.2(53.9) & 54 & 51.1 & 34.60 & 7.40 & 24.80 & 9.00\tabularnewline
 & LoRA8 & 42.1(34.8) & 31.5(27.2) & 37.2(33.5) & 59.6(48.4) & 53.4 & 53.1 & 31.80 & 8.80 & 34.90 & 12.20\tabularnewline
\midrule 
\multirow{3}{*}{concise2} & Full & 29.9(27.4) & 27.2(23.3) & 50.6(47.0) & 55.4(46.4) & 56.6 & 55.9 & 32.37 & 8.11 & 10.50 & 3.80\tabularnewline
 & LoRA64 & 30.5(28.0) & 32.5(27.8) & 44.5(39.0) & 63.2(53.6) & 54 & 51.2 & 35.50 & 8.10 & 21.80 & 7.70\tabularnewline
 & LoRA8 & 37.2(31.1) & 31.0(27.2) & 39.0(35.4) & 60.4(49.1) & 50.5 & 51.2 & 33.20 & 7.40 & 31.90 & 10.90\tabularnewline
\midrule 
\multirow{3}{*}{SemcoderGPT4o} & Full & 38.4(34.8) & 22.2(19.8) & 51.8(46.3) & 59.4(47.9) & 58.6 & 58 & 31.40 & 10.80 & 10.90 & 4.10\tabularnewline
 & LoRA64 & 37.2(32.9) & 26.7(24.1) & 42.7(38.4) & 59.9(50.1) & 54.8 & 50.4 & 34.40 & 8.80 & 23.10 & 8.60\tabularnewline
 & LoRA8 & 37.8(32.3) & 25.7(22.0) & 39.6(35.4) & 54.9(44.9) & 52.8 & 52.1 & 33.60 & 7.40 & 32.80 & 11.50\tabularnewline
\midrule 
\multirow{3}{*}{SemcoderGPT4o\_y} & Full & 31.7(28.7) & 31.5(28.0) & 51.8(48.2) & 62.7(50.1) & 59 & 60.2 & 30.60 & 7.40 & 14.70 & 5.60\tabularnewline
 & LoRA64 & 36.6(30.5) & 29.4(25.9) & 37.8(34.1) & 59.1(50.9) & 53.5 & 52.9 & 34.56 & 12.84 & 27.96 & 9.32\tabularnewline
 & LoRA8 & 39.0(33.5) & 26.7(23.5) & 37.2(32.3) & 55.9(44.9) & 52.9 & 51.9 & 32.20 & 5.40 & 33.60 & 11.80\tabularnewline
\midrule 
\multirow{3}{*}{Semcoder} & Full & 34.1(29.9) & 29.4(24.6) & 51.8(48.8) & 61.9(48.6) & 59.9 & 55.4 & 33.40 & 14.20 & 14.70 & 5.20\tabularnewline
 & LoRA64 & 28.7(25.6) & 31.5(27.2) & 42.7(36.6) & 59.9(50.4) & 54.9 & 50.4 & 34.70 & 6.80 & 24.80 & 8.70\tabularnewline
 & LoRA8 & 37.2(31.7) & 29.1(24.6) & 37.2(32.9) & 56.1(45.6) & 50.6 & 50.9 & 32.70 & 6.80 & 29.80 & 10.40\tabularnewline
\midrule 
\multirow{3}{*}{Semcoder\_y} & Full & 29.3(27.4) & 28.8(25.1) & 51.8(48.2) & 58.4(46.4) & 59 & 58.1 & 33.00 & 12.20 & 28.20 & 10.10\tabularnewline
 & LoRA64 & 33.5(29.3) & 27.8(24.6) & 42.1(36.6) & 55.6(44.9) & 54.1 & 53.9 & 35.61 & 10.81 & 26.52 & 8.75\tabularnewline
 & LoRA8 & 37.8(32.0) & 24.6(22.2) & 42.1(36.6) & 55.1(43.4) & 51.7 & 52 & 33.40 & 5.40 & 34.00 & 11.80\tabularnewline
\end{tabular}

}

\caption{A extend version of Table in LLaMA(llama3.1-8B) after finetuning with semantic information. }
\label{appindex_llama31_lora}
\end{table*}

\begin{table*}[t]
\small
\centering
\resizebox{0.95\textwidth}{!}
{

% Preview source code for paragraph 0
% Preview source code for paragraph 0

\begin{tabular}{cc|cc|cc|cc|cc|cc}
\multirow{2}{*}{} &  & \multicolumn{2}{c|}{repair} & \multicolumn{2}{c|}{NL2Code} & \multicolumn{2}{c|}{reasoning} & \multicolumn{2}{c|}{bigcodebench} & \multicolumn{2}{c}{Livecodebench}\tabularnewline
 &  & HE-R/(+) & MBPP-R/(+) & HE/(+) & MBPP/(+) & in\_predict & out\_predict & full & hard & easy pass@1 & overall pass@1\tabularnewline
\midrule 
Gemma2 &  & 37.2(33.5) & 20.9(19.6) & 40.2(34.1) & 63.7(51.9) & 49.2 & 41.5 & 29.80 & 6.80 & 32.80 & 11.40\tabularnewline
\midrule 
\multirow{3}{*}{only nl2code} & Full & 38.4(35.4) & 19.8(16.7) & 59.1(54.3) & 61.4(50.4) & 57.9 & 55.6 & 26.80 & 12.80 & 12.60 & 4.90\tabularnewline
 & LoRA64 & 50.0(42.1) & 24.9(22.8) & 54.9(47.0) & 69.4(55.1) & 57.2 & 60.1 & 37.81 & 11.49 & 35.84 & 12.95\tabularnewline
 & LoRA8 & 49.4(44.5) & 24.6(22.8) & 47.0(41.5) & 66.9(54.1) & 53.9 & 51.9 & 37.11 & 12.16 & 36.56 & 13.07\tabularnewline
\midrule 
\multirow{3}{*}{concise} & Full & 34.1(30.5) & 22.8(20.4) & 47.0(42.7) & 60.2(50.6) & 57.6 & 57.2 & 28.80 & 10.80 & 8.00 & 3.10\tabularnewline
 & LoRA64 & 35.4(31.7) & 33.1(28.3) & 50.0(45.1) & 66.9(52.6) & 59.8 & 51.2 & 38.86 & 10.81 & 26.88 & 9.09\tabularnewline
 & LoRA8 & 37.8(34.1) & 33.3(28.8) & 48.2(42.1) & 67.9(56.1) & 55.9 & 56.1 & 38.77 & 8.78 & 29.03 & 10.00\tabularnewline
\midrule 
\multirow{3}{*}{CodeExecutor} & Full & 28.7(25.6) & 22.8(19.3) & 42.1(37.2) & 59.4(48.9) & 58.9 & 58.2 & 27.30 & 13.50 & 8.80 & 3.10\tabularnewline
 & LoRA64 & 37.2(32.9) & 32.3(28.0) & 48.2(44.5) & 64.7(52.1) & 59.2 & 57.5 & 38.86 & 11.49 & 23.66 & 7.84\tabularnewline
 & LoRA8 & 36.6(31.7) & 33.9(29.1) & 45.1(39.6) & 66.2(55.9) & 51.5 & 49.2 & 37.63 & 8.11 & 25.81 & 9.20\tabularnewline
\midrule 
\multirow{3}{*}{w/o trace} & Full & 30.5(25.0) & 24.9(22.0) & 48.2(44.5) & 58.4(48.6) & 57.8 & 57.5 & 25.10 & 8.80 & 6.70 & 2.70\tabularnewline
 & LoRA64 & 31.7(28.7) & 29.9(26.7) & 48.2(42.7) & 66.9(54.6) & 58.8 & 56.4 & 38.77 & 11.49 & 23.30 & 8.30\tabularnewline
 & LoRA8 & 35.4(32.9) & 31.5(28.0) & 45.1(40.2) & 67.2(54.9) & 51.1 & 49 & 38.95 & 9.46 & 24.01 & 8.41\tabularnewline
\midrule 
\multirow{3}{*}{NeXT} & Full & 31.7(29.3) & 26.2(23.3) & 48.2(43.9) & 58.1(48.4) & 59.5 & 55.8 & 26.90 & 8.80 & 8.80 & 3.10\tabularnewline
 & LoRA64 & 38.4(34.8) & 32.0(28.0) & 47.6(42.1) & 66.2(53.6) & 59.5 & 54.5 & 38.42 & 12.16 & 28.32 & 9.32\tabularnewline
 & LoRA8 & 35.4(32.3) & 33.9(29.1) & 48.8(42.7) & 67.2(55.9) & 51.2 & 51 & 39.47 & 12.16 & 23.30 & 7.84\tabularnewline
\midrule 
\multirow{3}{*}{concise2} & Full & 26.2(24.4) & 25.1(21.4) & 48.2(43.9) & 56.9(47.6) & 58.9 & 56.2 & 29.00 & 7.40 & 8.00 & 2.90\tabularnewline
 & LoRA64 & 32.3(29.3) & 32.5(28.6) & 51.2(45.7) & 66.9(53.6) & 57.9 & 57.5 & 38.77 & 12.84 & 29.75 & 9.55\tabularnewline
 & LoRA8 & 40.2(34.8) & 31.7(27.8) & 44.5(37.8) & 67.4(55.1) & 46.5 & 51.7 & 38.86 & 6.76 & 29.03 & 10.00\tabularnewline
\midrule 
\multirow{3}{*}{SemcoderGPT4o} & Full & 35.4(31.1) & 24.1(21.7) & 51.8(48.8) & 62.9(50.1) & 58.9 & 56.5 & 29.50 & 7.40 & 8.40 & 2.90\tabularnewline
 & LoRA64 & 45.7(40.9) & 29.4(25.4) & 47.0(42.1) & 66.9(53.6) & 57.8 & 54.4 & 38.42 & 14.19 & 25.81 & 8.98\tabularnewline
 & LoRA8 & 45.1(39.0) & 28.6(24.9) & 47.6(42.7) & 67.7(55.1) & 41.1 & 52.5 & 38.95 & 10.14 & 26.16 & 9.32\tabularnewline
\midrule 
\multirow{3}{*}{SemcoderGPT4o\_y} & Full & 31.7(30.5) & 22.2(18.8) & 48.8(44.5) & 57.6(46.9) & 58.9 & 52.1 & 27.63 & 13.51 & 9.50 & 4.90\tabularnewline
 & LoRA64 & 32.3(29.9) & 29.1(25.9) & 47.0(40.2) & 68.7(55.9) & 59 & 57.6 & 39.04 & 10.81 & 30.11 & 10.11\tabularnewline
 & LoRA8 & 41.5(37.8) & 27.2(24.9) & 51.2(44.5) & 66.4(53.9) & 58.5 & 51.4 & 38.95 & 12.84 & 23.66 & 8.30\tabularnewline
\midrule 
\multirow{3}{*}{Semcoder} & Full & 31.1(27.4) & 26.2(22.2) & 53(50.0) & 62.2(53.1) & 58.9 & 56.8 & 27.60 & 13.50 & 13.00 & 4.80\tabularnewline
 & LoRA64 & 32.9(31.1) & 31.0(27.2) & 48.8(44.5) & 66.4(53.6) & 59.2 & 59.9 & 39.74 & 15.54 & 24.01 & 7.95\tabularnewline
 & LoRA8 & 39.6(35.4) & 29.9(26.2) & 43.9(38.4) & 66.2(55.6) & 56.9 & 55.4 & 39.74 & 10.14 & 21.86 & 7.61\tabularnewline
\midrule 
\multirow{3}{*}{Semcoder\_y} & Full & 26.8(24.4) & 24.3(20.9) & 49.4(45.7) & 59.4(48.9) & 58.5 & 52.1 & 23.25 & 5.41 & 17.00 & 5.80\tabularnewline
 & LoRA64 & 38.4(35.4) & 24.9(22.8) & 48.8(42.7) & 67.4(54.9) & 59.2 & 59.6 & 39.12 & 14.19 & 30.47 & 10.34\tabularnewline
 & LoRA8 & 40.2(33.5) & 24.6(22.2) & 43.3(37.2) & 67.9(56.1) & 51.1 & 48.5 & 38.77 & 13.51 & 27.96 & 10.00\tabularnewline
\end{tabular}

}

\caption{A extend version of Table in Gemma2(gemma2-9b) after finetuning with semantic information. }
\label{appindex_gemma_lora}
\end{table*}

\subsection{Prediction uncertainty analysis.}
 Finally, we explore the reason why fine-tuning with trace-based semantic information cannot enhance LLM's performance from the perspective of prediction uncertainty. We employ tasks and datasets provided by~\cite{uncertaintyye2024benchmarking} to measure the prediction uncertainty of fine-tuned models. Table~\ref{tab:uncertainty} presents the uncertainty scores and Table~\ref{tab:correlation} summarizes the correlation between prediction uncertainty and model performance. The correlation result demonstrates that for code generation tasks, there is a clear positive correlation between the uncertainty scores and the model performance. This indicates that the model is more certain about its prediction but has lower code generation capability. \textit{Takeaway:} Semantic information in fine-tuning datasets can truly boost the prediction confidence of Code LLMs, but fail to increase the prediction correctness. How to utilize semantic information to help fine-tune Code LLMs is still an open problem.

% %––––– Tables side-by-side –––––––––––––––––––––––––––––––––––––––––––––––––
% \begin{table*}[t]
%   \centering
%   % -------- left sub-table -------------------------------------------------
%   \subcaptionbox{Prediction uncertainty of LLMs.\label{tab:uncertainty}}
%     [.57\linewidth]% ← adjust width share if needed
%     {%
%       \resizebox{\linewidth}{!}{%
%       \begin{tabular}{lcccccccc}
%         & \textbf{Van.} & \textbf{NL2C} & \textbf{w/o T} & \textbf{Conc.} &
%         \textbf{CodeEx.} & \textbf{NExT} &
%         \textbf{\makecell{Sem\\(GPT4)}} & \textbf{Sem} \\ \toprule
%         \textbf{LLaMA}            & 1.90 & 3.66 & 2.43 & 2.98 & 2.95 & 2.74 & 2.44 & 2.32 \\
%         \textbf{DeepSeek-Coder}   & 3.25 & 3.74 & 3.60 & 3.47 & 3.74 & 3.49 & 3.62 & 3.47 \\
%         \textbf{Gemma2}           & 2.99 & 3.05 & 3.28 & 3.01 & 2.92 & 2.92 & 3.37 & 3.65 \\ \bottomrule
%       \end{tabular}}%
%     }
%   \hfill
%   % -------- right sub-table ------------------------------------------------
%   \subcaptionbox{Spearman rank correlation between prediction uncertainty and model performance.\label{tab:correlation}}
%     [.40\linewidth]%
%     {%
%       \resizebox{\linewidth}{!}{%
%       \vspace{-4cm}
%       \begin{tabular}{lcccccc}
%          & MBPP-R & MBPP & BigCode & LiveCB & CRUX & CRUX-O \\ \toprule
%         Coef.   & 0.356 & 0.759 & 0.440 & 0.357 & 0.598 & 0.228 \\
%         $p$-val & 0.088 & 1.7e-5 & 0.031 & 0.087 & 0.002 & 0.284 \\ \bottomrule
%       \end{tabular}}%
%     }

% \end{table*}

\begin{table*}[t]
  \centering
  \captionsetup[sub]{justification=centering, skip=4pt}
  % ---- left sub-table ------------------------------------------------------
  \begin{subtable}[t]{.70\textwidth}
    \centering
    \caption{Prediction uncertainty of LLMs.}
    \label{tab:uncertainty}
    \footnotesize
    \setlength{\tabcolsep}{3pt}
    \begin{tabular}{l*{8}{c}}
      \toprule
      & \textbf{Van.} & \textbf{NL2C} & \textbf{w/o T} & \textbf{Conc.} &
      \textbf{CodeEx.} & \textbf{NExT} &
      \textbf{\makecell{Sem\\(GPT4)}} & \textbf{Sem} \\
      \midrule
      \textbf{LLaMA}          & 1.90 & 3.66 & 2.43 & 2.98 & 2.95 & 2.74 & 2.44 & 2.32 \\
      \textbf{DeepSeek-Coder} & 3.25 & 3.74 & 3.60 & 3.47 & 3.74 & 3.49 & 3.62 & 3.47 \\
      \textbf{Gemma2}         & 2.99 & 3.05 & 3.28 & 3.01 & 2.92 & 2.92 & 3.37 & 3.65 \\
      \bottomrule
    \end{tabular}
  \end{subtable}
  \hspace{0.02\textwidth}
  % ---- right sub-table -----------------------------------------------------
  \begin{subtable}[t]{.53\textwidth}
    \centering
    \caption{Spearman rank correlation between\\prediction uncertainty and model performance.}
    \label{tab:correlation}
    \footnotesize
    \setlength{\tabcolsep}{2pt}
    \begin{tabular}{l*{6}{c}}
      \toprule
      & \textbf{\rotatebox{1}{MBPP-R}} & \textbf{\rotatebox{1}{MBPP}} & \textbf{\rotatebox{1}{BigCode}} & 
      \textbf{\rotatebox{1}{LiveCB}} & \textbf{\rotatebox{1}{CRUX}} & \textbf{\rotatebox{1}{CRUX-O}} \\
      \midrule
      \textbf{Coef.}   & 0.356 & 0.759 & 0.440 & 0.357 & 0.598 & 0.228 \\
      \textbf{$p$-val} & 0.088 & 1.7e-5 & 0.031 & 0.087 & 0.002 & 0.284 \\
      \bottomrule
    \end{tabular}
  \end{subtable}
\end{table*}

% \clearpage
\section {Experiment results of Test-scaling on MBPP-R }\label{more_rq3}

\begin{figure*}[t]
    \small
	\centering
	\includegraphics[width=0.85\textwidth]{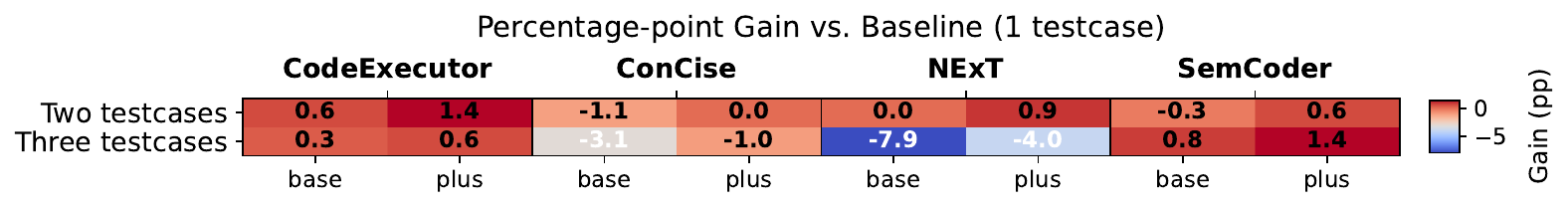}
	\caption{ Heat-map of percentage-point gains over the (one-test case) baseline. \textit{Rows} mark the aggregated test-case sizes. \textit{Columns} are grouped by trace representations each with its MBPP-R(base, plus) benchmark. Warm shades (reds) indicate positive gains.}
	\label{fig:effect_n}
\end{figure*}

Table~\ref{tbl_scaling_inference} summarizes the results of test-time scaling. It is clear that compared to open-source LLMs, closed-source LLMs perform significantly better at test time.
Comparison between inference with and without trace-based semantic information. First, we can see that, different from the findings from the fine-tuning investigation, inference with trace-based semantic information consistently boosts the performance of Code LLMs. In most cases~(98 out of 112), adding trace information enhances Code LLMs with an improvement by up to 10.85, a max improvement in NeXT with GPT-4o. This indicates that semantic information can guide LLMs in generating more correct code.
Comparison between different trace representations. Similar to previous findings, no single semantic representation consistently performs better than the others. SemCoder, which performs relatively better during fine-tuning, cannot stand out considering inference only. Concise, a variant of CodeExecutor designed by us performs the best under the instruction version of LLMs.    Different from fine-tuning, trace-based semantic information significantly enhances the performance of Code LLMs at test time. Sequential Revisions and Parallel are two optimal search strategies for test-time scaling.

\begin{table*}[t]
\centering
\resizebox{1.\textwidth}{!}
{

\begin{tabular}{l|c|c|c|c|c|c|c|c}
\multirow{2}{*}{} & \multicolumn{6}{c|}{Sequential Revisions} & \multirow{2}{*}{BoN} & \multirow{2}{*}{BeamSearch}\tabularnewline
 & \multicolumn{1}{c}{w/o trace(greedy)} & \multicolumn{1}{c}{Concise} & \multicolumn{1}{c}{CodeExecutor} & \multicolumn{1}{c}{NExT} & \multicolumn{1}{c}{SemCoder} & \multicolumn{1}{c|}{SemCoder(GPT4o)} &  & \tabularnewline
\midrule 
\textit{Close source Model}\\
GPT-4o & 50.79(44.71) & 60.05(50.79) & 59.79(50.79) & \underline{61.64(50.53)} & 58.73(50.0) & 59.41(51.10) & 54.46(42.23) & 61.29(51.71)\tabularnewline
deepseek-chat(V3)  & 53.17(47.88) & 61.11(54.23) & 60.85(52.38) & 61.11(53.7) & 61.11(52.4) & 61.31(52.68) & 56.37(57.45) & \underline{63.17(53.88)}\tabularnewline
\midrule 
\textit{Reasoning Compatible Model}\\
Marcon-o1 & 25.40(22.20) & 27.00(23.00) & 27.00(23.30) & 24.10(20.90) & 28.80(24.90) & 29.10(25.10) & 26.30(23.10) & \underline{29.40(25.10)}\tabularnewline
phi-4 & 39.15(33.86) & 43.65(38.10) & \underline{45.5(39.15)} & 42.86(38.10) & 44.71(39.68) & 44.71(40.12) & 41.23(36.45) & 45.15(41.86)\tabularnewline
\midrule 
\textit{Instruction version of Foundation Model}\\
CodeLlama-7b-Instruct-hf & 19.58(18.25) & \underline{20.11(18.52)} & 19.84(19.05) & 19.58(17.72) & 18.52(16.93) & 19.12(17.00) & 19.13(17.43) & 19.58(18.25)\tabularnewline
Llama-3.1-8B-Instruct & 28.84(27.51) & \underline{36.51(32.28)} & 32.01(29.89) & 32.01(28.57) & 33.86(31.48) & 34.21(32.14) & 29.21(28.43) & 33.54(31.42)\tabularnewline
deepseek-coder-6.7b-instruct & 24.87(23.54) & 23.81(22.49) & 25.66(23.54) & 23.54(22.22) & 30.16(27.51) & 30.56(28.31) & 25.87(24.34) & \underline{30.87(28.14)}\tabularnewline
\end{tabular}

}
\caption{\texttt{Pass@}1Comparing compute-optimal approaches on the code repair benchmark MBPP-R at test time, the numbers outside and inside parenthesis "()" indicate the base and plus versions of EvalPlus, respectively. w/o trace~(greedy) only interacts with the LLM via its initial (potentially buggy) code. In contrast, other sequential revision methods benefit from trace-based semantic information. The best results of EvalPlus' base highlights with \underline{underline} } 
\label{tbl_scaling_inference}
\end{table*}

% \begin{table}[t]
% \centering

% % Preview source code for paragraph 0

% \begin{tabular}{clccccc}
%  & round & 1 & 2 & 3 & 4 & 5\tabularnewline
% \hline 
%  & pass\_rate & 61.54 & 84.62 & 84.62 & 84.62 & 88.46\tabularnewline
% \hline 
% \multirow{5}{*}{error type} & extract-fail &  &  &  &  & \tabularnewline
%  & syntax-error & 50.96 & 9.13 & 6.25 & 3.85 & 4.33\tabularnewline
%  & execute-fail & 34.62 & 34.62 & 33.65 & 34.62 & 34.62\tabularnewline
%  & test-case-fail & 7.21 & 24.52 & 25.00 & 25.96 & 25.00\tabularnewline
%  & testcase-pass & 7.21 & 31.73 & 35.10 & 35.58 & 36.06\tabularnewline
% \end{tabular}

% \caption{\texttt{Pass@}1Comparing compute-optimal approaches on the code repair benchmark MBPP-R at test time, the numbers outside and inside parenthesis "()" indicate the base and plus versions of EvalPlus, respectively. w/o trace~(greedy) only interacts with the LLM via its initial (potentially buggy) code. In contrast, other sequential revision methods benefit from trace-based semantic information. The best results of EvalPlus' base highlights with \underline{underline} } 
% \label{tbl_error_type }
% \end{table}

%––––– Tables side-by-side –––––––––––––––––––––––––––––––––––––––––––––––––
\begin{table*}[t]
  \centering
  % -------- left sub-table -------------------------------------------------

  \subcaptionbox{Pass-rate improvement and error-type distribution over five self-debugging rounds. The overall pass rate climbs from 61.54 \% in Round 1 to 88.46 \% by Round 5, while syntax errors drop sharply and an increasing share of examples transitions into the \textit{testcase-pass} category. Percentages are shown for each round; blank cells indicate zero occurrences..\label{tab:errtype}}
    [.50\linewidth]% ← adjust width share if needed
    {%
      \resizebox{\linewidth}{!}{%

\begin{tabular}{clccccc}
\toprule
 & Round & 1 & 2 & 3 & 4 & 5\tabularnewline
\midrule 
 & Pass@1 & 61.54 & 84.62 & 84.62 & 84.62 & 88.46\tabularnewline
\midrule 
\multirow{5}{*} & SyntaxError & 50.96 & 9.13 & 6.25 & 3.85 & 4.33\tabularnewline
 & ExecuteFail & 34.62 & 34.62 & 33.65 & 34.62 & 34.62\tabularnewline
 & TestcaseFail & 7.21 & 24.52 & 25.00 & 25.96 & 25.00\tabularnewline
\bottomrule
\end{tabular}

      }%
    }
  \hfill
  % -------- right sub-table ------------------------------------------------
  % \subcaptionbox{Spearman rank correlation between prediction uncertainty and model performance.\label{tab:correlation}}
    % [.40\linewidth]%
    {%

    }

\end{table*}

% \clearpage
\section{More Prompt Example~\label{full_example}}

\subsection{Prompt templates}
\label{sec:appendix_prompts}
% We also provide detailed prompts used in our experiments in \ref{fig:self-refine} to \ref{fig:generation-prompt}. These prompts are generated automatically by DSPy~\cite{khattab2024dspy}.

We also provide detailed prompts used in our experiments in website. Please refer to our github repository~\footnote{\url{https://github.com/tracewise-probing/tracewise_probing}} for full details.